\documentstyle[amssymb,11pt,epsfig,psfig]{article}
1.60cm \makeatletter \@addtoreset{equation}{section}

\makeatother
\def\beq{\begin{equation}}
\def\eeq{\end{equation}}

\def\text#1{\mbox{\scriptsize #1}}

\begin{document}

\noindent{\Large {\bf Casimir effect for scalar fields under Robin
boundary conditions on plates}}

\vspace*{5mm} \noindent August Romeo$^{a,}$\footnote{%
E-mail: romeo@ieec.fcr.es} and Aram A. Saharian$^{b,}$\footnote{%
E-mail: saharyan@server.physdep.r.am} \newline \noindent ${}^a$
Institut d'Estudis Espacials de Catalunya (IEEC/CSIC), Institut de
Ci{\`e}ncies de l'Espai (CSIC), \newline Edifici Nexus-201 - c.
Gran Capit\`a 2-4, 08034 Barcelona \newline \noindent ${}^b$
Department of Physics, Yerevan State University, 1 Alex Manoogian
St, 375049 Yerevan, Armenia \vspace*{10mm}

\noindent {\bf Abstract.} We study the Casimir effect for scalar
fields with general curvature coupling subject to mixed boundary
conditions $(1+\beta _{m}n^{\mu }\partial _{\mu })\varphi =0$ at
$x=a_{m}$ on one ($m=1$) and two ($m=1,2$) parallel plates at a
distance $a\equiv a_{2}-a_{1}$ from each other. Making use of the
generalized Abel-Plana formula previously established by one of
the authors
\cite{Sahrev}, the Casimir energy densities are obtained as functions of $%
\beta _{1}$ and of $\beta _{1}$,$\beta _{2}$,$a$, respectively. In
the case of two parallel plates, a decomposition of the total
Casimir energy into volumic and superficial contributions is
provided. The possibility of finding a vanishing energy for
particular parameter choices is shown, and the existence of a
minimum to the surface part is also observed. We show that there
is a region in the space of parameters defining the boundary
conditions in which the Casimir forces are repulsive for small
distances and attractive for large distances. This yields to an
interesting possibility for stabilizing the distance between the
plates by using the vacuum forces.

\vspace*{5mm}

\section{Introduction}

Although the existing literature about the Casimir effect is quite
sizable in volume (for reviews see. e.g. \cite{PMG}), we feel that
relatively little
attention has been devoted to quantum fields subject to Robin ---or {\it %
mixed}--- boundary conditions on plates. A possible reason is that
this type of condition appears when decomposing the modes of the
electromagnetic field in the presence of perfectly conducting
spheres (see refs. \cite{sph}-\cite {NP}), but are not required in
the analogous problem with parallel plates, where the mode set can
be divided into eigenmodes satisfying Dirichlet and Neumann
conditions separately.

However, Robin conditions can be made conformally invariant, while
purely-Neumann ones cannot. Thus, Robin-type conditions are needed
when one deals with conformally invariant theories in the presence
of boundaries and wishes to preserve this invariance. The
importance of conformal invariance in problems related to the
Casimir effect has been emphasized, e.g. in refs. \cite{ER,KCD}
(see also \cite{Can}). The Casimir energy-momentum tensor on
background of conformally flat geometries can be obtained by the
standard transformation from the corresponding flat spacetime
result (see \cite{Birrel}). For instance, by this way in
\cite{SetSah} the Casimir effect for a scalar field with Dirichlet
boundary condition is investigated on background of a static
domain wall geometry. To derive the Casimir characteristics via
conformal transformations for the case of Neumann boundary
condition we need to have the corresponding flat spacetime
counterpart with Robin boundary conditions and Robin coefficient
related to the conformal factor. It is interesting to note that
the quantum scalar field satisfying Robin condition on the
boundary of cavity violates the Bekenstein's entropy-to-energy
bound near certain points in the space of the parameter defining
the boundary condition \cite{Solod}. The Robin boundary conditions
are an extension of the ones imposed on perfectly conducting
boundaries and may, in some geometries, be useful for depicting
the finite penetration of the field into the boundary with the
"skin-depth" parameter related to the Robin coefficient
\cite{Mostep}. On the other hand, the relevance of mixed-type
boundary conditions to spacetime models and quantum gravity has
been highlighted in refs. \cite{Mo}, \cite{EK}. This type of
conditions naturally arises for the scalar and fermion bulk fields
in the Randall-Sundrum model \cite{Gherg}. In this model the bulk
geometry is a slice of anti-de Sitter space and the corresponding
Robin coefficient is related to the curvature scale of this space.

In the present work we discuss several aspects of the Casimir
effect for a massless scalar field, with curvature coupling,
obeying Robin boundary conditions on one or two parallel plates on
background of a flat ${\cal D}$-dimensional spacetime. The
dimensional dependence of physical quantities in the Casimir
effect is of considerable interest and is investigated for various
types of geometries (see, for instance, \cite{DDep},
\cite{Sahsph}, \cite{Sahcyl}). In sec. \ref{sec:1conf} we explain
how Robin conditions can adopt a conformally invariant form. The
Casimir effect with one plane boundary is considered in sec.
\ref{sec:1pl}, while sec. \ref{sec:2pl} is dedicated to the set-up
where two parallel plates are present. Then, the volume and
surface contributions to the total Casimir energy (for this second
case) are analyzed in sec. \ref{sec:totcas}. Our ending comments
follow in sec. \ref{sec:endcom}.

\section{Conformal invariance and boundary conditions} \label{sec:1conf}

Let's consider a massless scalar field $\varphi $ with curvature coupling $%
\xi $ on background of a ${\cal D}$-dimensional spacetime manifold
$M$ with boundary $\partial M$. The action for this field is
\begin{equation}
S[\varphi ,g]=-\frac{1}{2}\int_{M}d^{{\cal
D}}x\,\sqrt{-g}\,\varphi \left[ \Box +\xi R\right] \varphi \,,
\label{action}
\end{equation}
where $\Box $ - is the Laplace-Beltrami operator. The Lagrangian
corresponding to (\ref{action}) differs from the often used
Lagrangian by a
total divergence leading to the additional surface term $\frac{1}{2}%
\int_{\partial M}d^{{\cal D}-1}x\,\sqrt{-g}\,n^{\mu }\varphi
\partial _{\mu }\varphi $ with $n^{\mu }$ being the unit normal
vector to $\partial M$. As it has been noted in ref.\cite{KCD},
this term plays a crucial role in the cancellations between
surface and volume divergences. Note that the additional surface
term is zero for Dirichlet and Neumann boundary conditions on
$\partial M$, but is nonzero for the more general Robin case.

Consider a conformal transformation realized by a Weyl rescaling
of the spacetime metric
\begin{equation}
g_{\mu \nu }(x)\longrightarrow \Omega ^{2}(x)g_{\mu \nu }(x).
\label{transfg}
\end{equation}
Under these transformations, the $\varphi $ field will change by a
rule of the type
\begin{equation}
\varphi (x)\longrightarrow \Omega ^{\alpha }(x)\varphi (x).
\label{transfphi1}
\end{equation}
As a result, the action undergoes the following transformation:
\begin{eqnarray}
S[\Omega ^{\alpha }\varphi ,\Omega ^{2}g]&=&-\frac{1}{2}\int d^{%
{\cal D}}x\,\sqrt{-g}\,\Omega ^{{\cal D}-2+2\alpha } \left\{
\varphi \left[ \Box +\zeta R\right] \varphi +[\alpha +2\xi ({\cal D}-1)]%
\frac{\Box \Omega }{\Omega }\varphi ^{2}\right. \nonumber \\
& & +(2\alpha +{\cal D}-2)g^{\mu \nu }{\frac{\,\partial _{\mu
}\Omega }{\Omega }}\,\varphi \partial _{\nu }\varphi \nonumber \\
& & \left. +[\alpha (\alpha +{\cal D}-3)-\xi ({\cal D}-1)({\cal D}%
-4)]\,g^{\mu \nu }{\frac{\,\partial _{\mu }\Omega \,}{\Omega }}\,\frac{%
\,\partial _{\nu }\Omega \,}{\Omega }\varphi ^{2}\right\} .
\label{acconf}
\end{eqnarray}
The action $S$ will be invariant if ${\cal D}-2+2\alpha =0$ and
all the terms containing derivatives of $\Omega $ vanish. These
two requirements are satisfied provided that
\begin{equation}
\left\{
\begin{array}{lll}
\alpha & = & \displaystyle-{\frac{{\cal D}-2}{2}}, \\
\xi & = & \displaystyle{\frac{{\cal D}-2}{4({\cal D}-1)}}\equiv
\xi _{c}.
\end{array}
\right.  \label{transfphi2}
\end{equation}
Next, we shall consider the effect of the transformation on a
boundary condition of the Neumann type
\begin{equation}
n^{\mu }\,\nabla _{\mu }\varphi (x)=0,  \label{bcN}
\end{equation}
where $n^\mu $ is a normal space-like vector (i.e., $g_{\mu \nu }{n}^{\mu }{n}%
^{\nu }=-1$) perpendicular to the boundary, and covariant
derivative $\nabla
_{\mu }$ reduces, in this case, to the ordinary partial derivative because $%
\varphi $ is just a scalar function. Let $\overline{n}^\mu $
denotes the transformed version of $n^\mu $. If we require that
the normalization be maintained, we shall have $\displaystyle
\Omega ^{2}g_{\mu \nu }\overline{n}^{\mu }\overline{n}^{\nu }=-1,$
whose solution is
\begin{equation}
\overline{n}^{\mu }={\frac{1}{\Omega }}n^{\mu }.  \label{transfn}
\end{equation}
Taking into account (\ref{transfg}), (\ref{transfphi1}),
(\ref{transfphi2})
and (\ref{transfn}), we realize that the l.h.s of the boundary condition (%
\ref{bcN}) transforms as
\begin{equation}
n^{\mu }\nabla _{\mu }\varphi \longrightarrow \Omega ^{-{\frac{{\cal D}-1}{2}%
}}\left( n^{\mu }\nabla _{\mu }\varphi -{\frac{{\cal
D}-2}{2}}{\frac{n^{\mu }\partial _{\mu }\Omega }{\Omega }}\varphi
\right) .
\end{equation}
The presence of the second term indicates that a boundary
condition of purely-Neumann type cannot be maintained under
general conformal transformations.

Similarly, if, instead of (\ref{bcN}), one takes a generic Robin
boundary condition
\begin{equation}
\left( \Psi(x) + n^{\mu} \, \nabla_{\mu} \right) \varphi(x) = 0,
\label{bcR}
\end{equation}
one can readily observe that it changes according to the rule
\begin{equation}
\left( \Psi + n^{\mu} \, \nabla_{\mu} \right) \varphi
\longrightarrow \Omega^{-{\frac{{\cal D}-1 }{2}}} \left( \Omega
\overline{\Psi} \varphi +n^{\mu} \nabla_{\mu} \varphi
-{\frac{{\cal D}-2 }{2}} {\frac{ n^{\mu}
\partial_{\mu} \Omega }{\Omega }} \varphi \right) ,  \label{transfR1}
\end{equation}
where $\overline{\Psi}$ indicates the result of transforming the
$\Psi$ function. The boundary condition (\ref{bcR}) can be
preserved only if the transformed version is proportional to the
initial form. Thus, one demands
that the r.h.s of (\ref{transfR1}) be equal to $\Omega^{-{\frac{{\cal D}-1 }{%
2}}} \left( \Psi + n^{\mu}\nabla_{\mu} \right) \varphi$. This
leads to a specific transformation rule for the $\Psi$ function,
which reads
\begin{equation}
\overline{\Psi}= \Omega^{-1} \left( \Psi + {\frac{{\cal D}-2 }{2}}
{\frac{ n^{\mu} \partial_{\mu} \Omega }{\Omega }} \right) ,
\label{transfPsi}
\end{equation}
as already observed in ref.\cite{KCD}.

Now, suppose that we have a valid $\Psi $ function satisfying
(\ref {transfPsi}). We can consider $n$ along the $x$-axis and set
boundary conditions on the planes $x=a_{1}$ and $x=a_{2}$.
Provided that $\Psi (x=a_{1})\neq 0$ and $\Psi (x=a_{2})\neq 0$,
we may write the boundary conditions at these points in the form
\begin{equation}
\begin{array}{ll}
(1+\beta _{m}(-1)^{m-1}\partial _{x})\varphi =0 & \mbox{at $x=a_m$,}%
\hspace*{5mm}\mbox{ $m=1,2$,}
\end{array}
\label{Rbcx0a}
\end{equation}
where
\begin{equation}
\beta _{m}={\frac{1}{\Psi (x=a_{m})}},\ \mbox{$m=1,2$.}
\label{defsbt1bt2}
\end{equation}

One may consider the subgroup of transformations in which $\Omega
$ does not
depend on the $x$-coordinate. Then, a possible $\Psi $ is given by $%
\displaystyle\Psi \lbrack (g)]=(-g)^{-1/2{\cal D}}$. These
particular transformations correspond to the restriction of the
initial group to planes parallel to the plates. In a strictly
Euclidean or Minkowskian spacetime, this form of $\Psi $ would
imply $\beta _{2}=\beta _{1}$.

\section{Casimir stresses for a single plate geometry} \label{sec:1pl}

In this section we will consider scalar field in $D$-spatial
dimensions
---thus, $D={\cal D}-1$--- with general coupling $\zeta $ satisfying Robin
boundary condition on the single boundary $x=0$ on background of a
flat spacetime. Such a situation is like limiting
eq.(\ref{Rbcx0a}) to $m=1$ only, and with $a_{1}=0$, i.e.,
\begin{equation}
(1+\beta _{1}n^{\mu }\partial _{\mu })\varphi (t,{\bf x})=(1+\beta
_{1}\partial _{x})\varphi (t,{\bf x})=0,\quad x=0.  \label{Robin1plate}
\end{equation}
Here we consider the vacuum fluctuations in the region $x\geq 0$. For the
region $x\leq 0$ the boundary condition has the form $(1-\beta _{1}\partial
_{x})\varphi (t,{\bf x})=0$ at $x=0$. The corresponding results can be
obtained from the previous case replacing $\beta _{1}\rightarrow -\beta _{1}$%
.

The eigenfunctions satisfying boundary condition (\ref{Robin1plate}) are in
form
\begin{equation}
\varphi _{{\bf k}}(t,{\bf x})=\frac{e^{i{\bf k}_{\perp }{\bf x}-i\omega t}}{%
\sqrt{2^{D-1}\pi ^{D}\omega }}\cos (kx+\alpha _{1}),  \label{eigfunc1plate}
\end{equation}
where ${\bf k}=(k,{\bf k}_{\perp })$, $\omega =|{\bf k}|$, $0\leq k<\infty $%
, and
\begin{equation}
\sin \alpha _{1}=\frac{1}{\sqrt{k^{2}\beta _{1}^{2}+1}},\quad \cos \alpha
_{1}=\frac{k\beta _{1}}{\sqrt{k^{2}\beta _{1}^{2}+1}}.  \label{alf01plate}
\end{equation}
In the case $\beta _{1}>0$ there is also a purely imaginary eigenvalue $%
k=i/\beta _{1}$ with the normalized eigenfunction
\begin{equation}
\varphi _{{\bf k}_{\perp }}^{{\rm (im)}}(t,{\bf x})=\frac{e^{i{\bf k}_{\perp
}{\bf x}-i\omega t-x/\beta _{1}}}{\sqrt{(2\pi )^{D-1}\omega \beta _{1}}}%
,\quad \omega =\sqrt{k_{\perp }^{2}-1/\beta _{1}^{2}},  \label{imeig}
\end{equation}
where $k_{\perp }=\left| {\bf k}_{\perp }\right| \geq 1/\beta _{1}$.

\subsection{Vacuum densities} \label{sec:1pl1}

From the symmetry of the problem it follows that the vacuum expectation
values (v.e.v.) for the energy-momentum tensor (EMT) have the form
\begin{equation}
\langle 0|T_{i}^{k}|0\rangle ={\rm diag}(\varepsilon ,-p,-p_{\perp },\ldots
,-p_{\perp }).  \label{emtdiag}
\end{equation}
The corresponding energy density $\varepsilon $ and effective pressures $p$,
$p_{\perp }$ can be derived by evaluating the mode sum
\begin{equation}
\langle 0|T_{ik}(x)|0\rangle =\sum_{\alpha }T_{ik}\left\{ \varphi _{\alpha
}(x),\varphi _{\alpha }^{\ast }(x)\right\} ,  \label{vevEMT}
\end{equation}
where $\varphi _{\alpha }(x)=\left( \varphi _{{\bf k}},\varphi _{{\bf k}%
_{\perp }}^{{\rm (im)}}\right) $ is a complete set of solutions to the field
equation and the bilinear form $T_{ik}\{f,g\}$ is determined by classical
energy-momentum tensor for the scalar field (see, e.g., \cite{Birrel}).
Using the field equation we will present it in the form
\begin{equation}
T_{ik}=\partial _{i}\varphi \partial _{k}\varphi +\left[ (\xi -\frac{1}{4}%
)g_{ik}\Box -\xi \partial _{i}\partial _{k}\right] \varphi ^{2}  \label{EMT}
\end{equation}
Using formula (\ref{vevEMT}) with eigenfunctions (\ref{eigfunc1plate}), (\ref
{imeig}) and EMT from (\ref{EMT}) one finds (no summation over $i$)
\begin{eqnarray}
\langle 0|T_{ik}(x)|0\rangle  &=&\langle 0_{M}|T_{ik}(x)|0_{M}\rangle
+\delta _{ik}\int \frac{d^{D-1}{\bf k}_{\perp }}{(2\pi )^{D}}%
\int_{0}^{\infty }dk\frac{A_{i}(k)}{\omega }\cos \left( 2kx+2\alpha
_{1}\right) + \nonumber  \\
&&+\delta _{ik}\theta (\beta _{1})
\frac{e^{-2x/\beta _{1}}}{2\beta _{1}}\int \frac{d^{D-1}{\bf k}_{\perp }}{%
(2\pi )^{D-1}}\frac{A_{i}(i/\beta _{1})}{\sqrt{k_{\perp
}^{2}-1/\beta _{1}^{2}}},  \label{EMT1}
\end{eqnarray}
where $\theta (x)$ stands for Heaviside step function,
\begin{equation}
A_{0}(k)=k_{\perp }^{2}+4\xi k^{2},\quad A_{1}(k)=0,\quad A_{i}(k)=\frac{%
k_{\perp }^{2}}{D-1}-(4\xi -1)k^{2},\quad i=2,...,D,  \label{Ai}
\end{equation}
and
\begin{equation}
\langle 0_{M}|T_{ik}(x)|0_{M}\rangle =\int \frac{d^{D-1}{\bf k}_{\perp }}{%
(2\pi )^{D}}\int_{0}^{\infty }\frac{dk}{\omega }\,{\rm diag}\left(
\omega ^{2},k^{2},\frac{k_{\perp }^{2}}{D-1},\ldots
,\frac{k_{\perp }^{2}}{D-1}\right) \label{EMTMink}
\end{equation}
are the corresponding quantities for the Minkowski vacuum
$|0_{M}\rangle $. The last summand on the right of (\ref{EMT1})
corresponds to the contribution of eigenfunctions (\ref{imeig}).
In this term the integration over ${\bf k}_{\perp }$ goes in the
region $\left| {\bf k}_{\perp }\right| \geq 1/\beta _{1}$.

In eq. (\ref{EMT1}) the integrals over ${\bf k}_{\perp }$ may be
evaluated by using the formulae
\begin{eqnarray}
\int d^{D-1}{\bf k}_{\perp }g(k_{\perp })&=&\frac{2\pi ^{(D-1)/2}}{\Gamma (%
\frac{D-1}{2})}\int_{0}^{\infty }k_{\perp }^{D-2}g(k_{\perp
})dk_{\perp } \label{int1} \\
\int_{0}^{\infty }\frac{y^{n}dy}{\sqrt{y^{2}+1}}&=&\frac{1}{2\sqrt{\pi }}%
\Gamma \left( -\frac{n}{2}\right) \Gamma \left(
\frac{n+1}{2}\right) \label{int2} \\
\int_{1}^{\infty }\frac{y^{n}dy}{\sqrt{y^{2}-1}}&=&\frac{1}{2\sqrt{\pi }}%
\Gamma \left( -\frac{n}{2}\right) \Gamma \left(
\frac{n+1}{2}\right) \cos \frac{\pi n}{2}.  \label{int2im}
\end{eqnarray}
As a result for the difference between v.e.v. for vacuums $|0\rangle $ and $%
|0_{M}\rangle $ one obtains
\begin{eqnarray}
\left\langle T_{ik}\right\rangle _{SUB} &=&\langle 0|T_{ik}(x)|0\rangle
-\langle 0_{M}|T_{ik}(x)|0_{M}\rangle =(\xi _{c}-\xi )\frac{D\Gamma (-D/2)}{%
2^{D-1}\pi ^{D/2+1}}{\rm diag}(1,0,-1,\ldots ,-1)   \nonumber\\
&&\times \left[ \int_{0}^{\infty }dkk^{D}\cos \left( 2kx+2\alpha
_{1}(k)\right) +2\pi \theta (\beta _{1})\frac{e^{-2x/\beta
_{1}}}{\beta _{1}^{D+1}}\cos \frac{\pi D}{2}\right] ,
\label{reg1plate}
\end{eqnarray}
where $\xi =\xi _{c}\equiv (D-1)/4D$ corresponds to the
conformally coupled scalar field (recall eq. (\ref{transfphi2})
with ${\cal D}=D+1$) and the function $\alpha _{1}(k)$ is defined
as (\ref{alf01plate}). Using these relations the integral over $k$
can be presented in the form
\begin{equation}
\int_{0}^{\infty }dkk^{D}\cos \left( 2kx+2\alpha _{1}(k)\right)
=\int_{0}^{\infty }dkk^{D}\cos \left( 2kx\right) -2\int_{0}^{\infty }dkk^{D}%
\frac{\cos \left( 2kx\right) +k\beta _{1}\sin (2kx)}{k^{2}\beta _{1}^{2}+1}.
\label{intcos2}
\end{equation}
To evaluate the second integral on the right we will use the formula \cite
{Prudnikov}
\begin{equation}
\int_{0}^{\infty }\frac{k^{n-1}e^{-p_{0}k}}{k^{2}+z^{2}}dk=-\frac{z^{n-2}}{2}%
\Gamma (n-1)\left[ e^{ip_{0}z+\pi ni/2}\Gamma (2-n,ip_{0}z)+e^{-ip_{0}z-\pi
ni/2}\Gamma (2-n,-ip_{0}z)\right] ,  \label{int3}
\end{equation}
where $\mbox{\rm Re}\,(z),\mbox{\rm Re}\,(n),\mbox{\rm Re}\,(p_{0})>0$, and $%
\Gamma (a,z)$ is the incomplete gamma-function (see, e.g., \cite{Abramowiz}%
). The corresponding integrals in (\ref{intcos2}) with $\cos $ and $\sin $
can be obtained as real and imaginary parts with $p_{0}=\alpha -2ix$, $%
\alpha \rightarrow +0$. Using the formulae \cite{Abramowiz}
\begin{equation}
\Gamma (-n,y)=\frac{(-1)^{n}}{n!}\left[ E_{1}(y)-e^{-y}\sum_{j=0}^{n-1}\frac{%
(-1)^{j}j!}{y^{j+1}}\right]   \label{GamE1}
\end{equation}
\begin{equation}
E_{1}(-y\pm i0)=-\mbox{\rm Ei}(y)\mp i\pi   \label{E1Ei}
\end{equation}
the corresponding v.e.v. can be presented in the terms of the functions $%
E_{1}(y)$ and $\mbox{\rm Ei}(y)$. As a result it can be seen that
the difference between v.e.v. on the left of (\ref{reg1plate}%
) is finite for $x>0$ and can be presented in the form

\begin{eqnarray}
\left\langle T_{00}\right\rangle _{SUB} & = & -\frac{4D(\xi -\xi
_{c})}{\pi ^{D/2}\Gamma (D/2+1)(4\left| x\right| )^{D+1}}\left\{
\Gamma \left( D+1\right) +y^{D+1}\left[ e^{\left| y\right|
}E_{1}(\left| y\right|
)\left( 1-\frac{\left| y\right| }{y}\right) \right. \right.  \nonumber \\
& & \left. \left. -e^{-\left| y\right| }\mbox{\rm Ei}(\left|
y\right| )\left( 1+\frac{\left| y\right| }{y}\right) +2\sum_{j=0}^{D-1}\frac{%
j!}{y^{j+1}}\right] \right\} ,\quad y=2x/\beta _{1},
\label{eps1plate}
\end{eqnarray}
and
\begin{equation}
\begin{array}{rcl}
\left\langle T_{ii}\right\rangle _{SUB} & = & -\left\langle
T_{00}\right\rangle _{SUB},\quad i=2,\ldots ,D \\
\left\langle T_{11}\right\rangle _{SUB} & = & 0.
\end{array}
\label{regp1plate}
\end{equation}
Note that in these formulae the term in the v.e.v.
(\ref{reg1plate}) coming from the contribution of the
eigenfunctions (\ref{imeig}) (and divergent for even values of
$D$) is cancelled with the corresponding term coming from the
integral (\ref{intcos2}). As a result the subtracted v.e.v. are
finite for all values $y>0$.

We have considered the region $x>0$. As it have been mentioned above the
corresponding formulae for $x<0$ can be obtained from (\ref{eps1plate})
replacing $\beta _{1}\rightarrow -\beta _{1}$. As a result the value of $y$
remains the same and the distribution for v.e.v. (\ref{eps1plate}) is
symmetric for regions $x>0$ and $x<0$ . The cases for the Dirichlet and
Neumann boundary conditions are obtained from (\ref{eps1plate}) in limits $%
\beta _{1}\rightarrow 0$ and $\beta _{1}\rightarrow \infty $, respectively:
\begin{equation}
(\left\langle T_{00}\right\rangle _{SUB})_{\mbox{\scriptsize Dirichlet}%
}=-(\left\langle T_{00}\right\rangle _{SUB})_{\mbox{\scriptsize Neumann}}=%
\frac{D(\xi -\xi _{c})}{2^{D}\pi ^{(D+1)/2}\left| x\right| ^{D+1}}\Gamma
\left( \frac{D+1}{2}\right) .  \label{Dirichlet1plate}
\end{equation}
Note that the first term in (\ref{eps1plate}) coincides with the energy
density corresponding to Neumann case. As follows from (\ref{eps1plate}),
the regularized v.e.v. is zero for a conformally coupled field. This result
can be obtained also without explicit calculations by using the continuity
equation and zero trace condition for the EMT.

Using the asymptotic formulae for the functions $E_{1}(z)$ and $\mbox{\rm Ei}%
(z)$ from (\ref{eps1plate}) one obtains the following asymptotic expansion
of the vacuum energy density for large $y$:
\begin{equation}
\left\langle T_{00}\right\rangle _{SUB}=(\left\langle T_{00}\right\rangle
_{SUB})_{_{\mbox{\scriptsize Dirichlet}}}+\frac{D(\xi -\xi _{c})}{%
2^{2D-1}\pi ^{(D+1)/2}\left| x\right| ^{D+1}}\sum_{j=D+1}^{\infty }\frac{j!}{%
y^{j-D}}.  \label{farfrom1pl}
\end{equation}
As we see at distances far from the plate the vacuum energy density
coincides with that for the Dirichlet case and is positive for $\xi >\xi
_{c} $ and is negative for $\xi <\xi _{c}$. At distances near the plate , $%
\left| y\right| \ll 1$ the vacuum energy density is dominated by the first
summand in the figure brackets in (\ref{eps1plate}). As it have been noted
this summand coincides with the energy density for the Neumann case and
hence has opposite sign compared to the case of far distances. As a result
the vacuum energy density has a positive maximum or negative minimum
(depending on the sign of $\xi -\xi _{c}$) for some intermediate value of $x$%
. Near the plate from (\ref{eps1plate}) we have the following expansion for
the vacuum energy density
\begin{equation}
\left\langle T_{00}\right\rangle _{SUB}=(\left\langle T_{00}\right\rangle
_{SUB})_{_{\mbox{\scriptsize Neumann}}}\left[ 1+\frac{2}{D!}\left(
\sum_{j=0}^{D-1}j!\,y^{D-j}-y^{D+1}\ln \left| y\right| \right) \right] +%
{\cal O}(1).  \label{near1plate}
\end{equation}
All terms presented lead to the diverging contribution to the energy density
at the plate surface. These surface divergences are well known in quantum
field theory with boundaries and are investigated near arbitrary shaped
smooth boundary for the minimal and conformal scalar and electromagnetic
fields \cite{Can},\cite{KCD}.

\begin{figure}[tbph]
\begin{center}
\begin{tabular}{cc}
\epsfig{figure=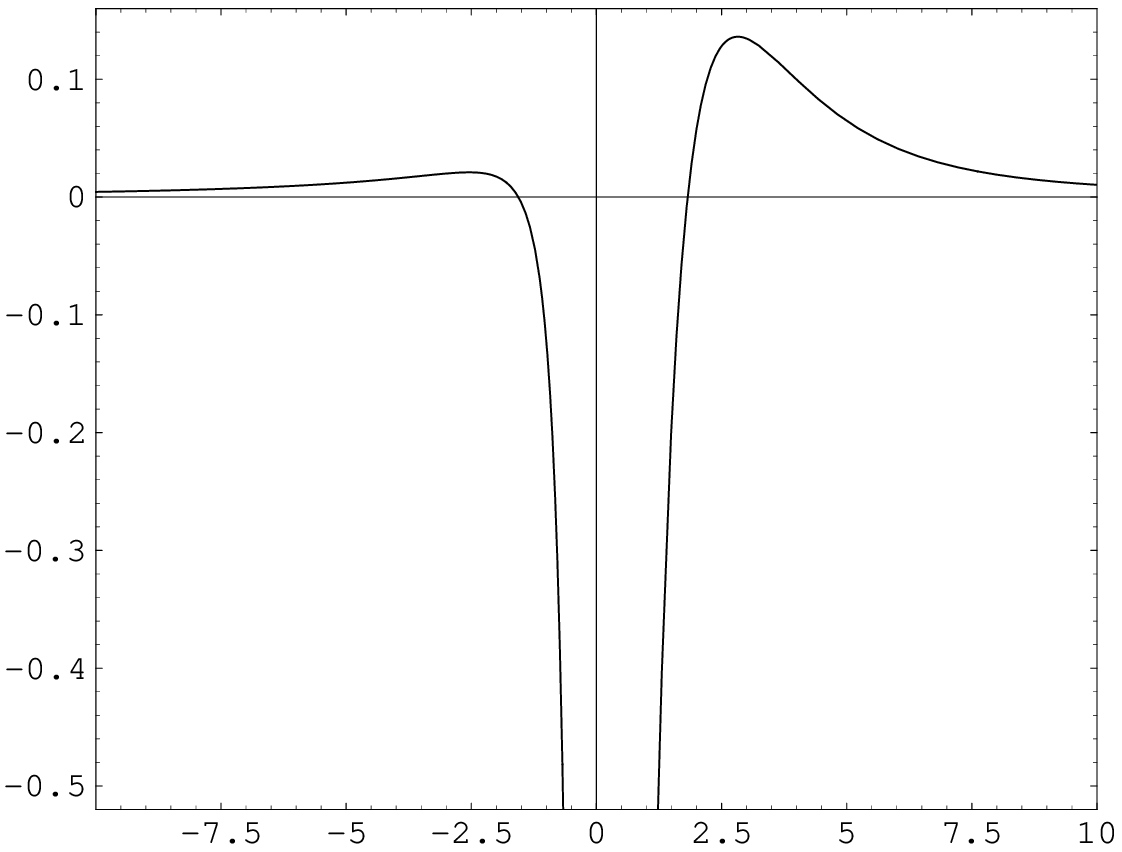,width=5.8cm,height=5.8cm} & %
\epsfig{figure=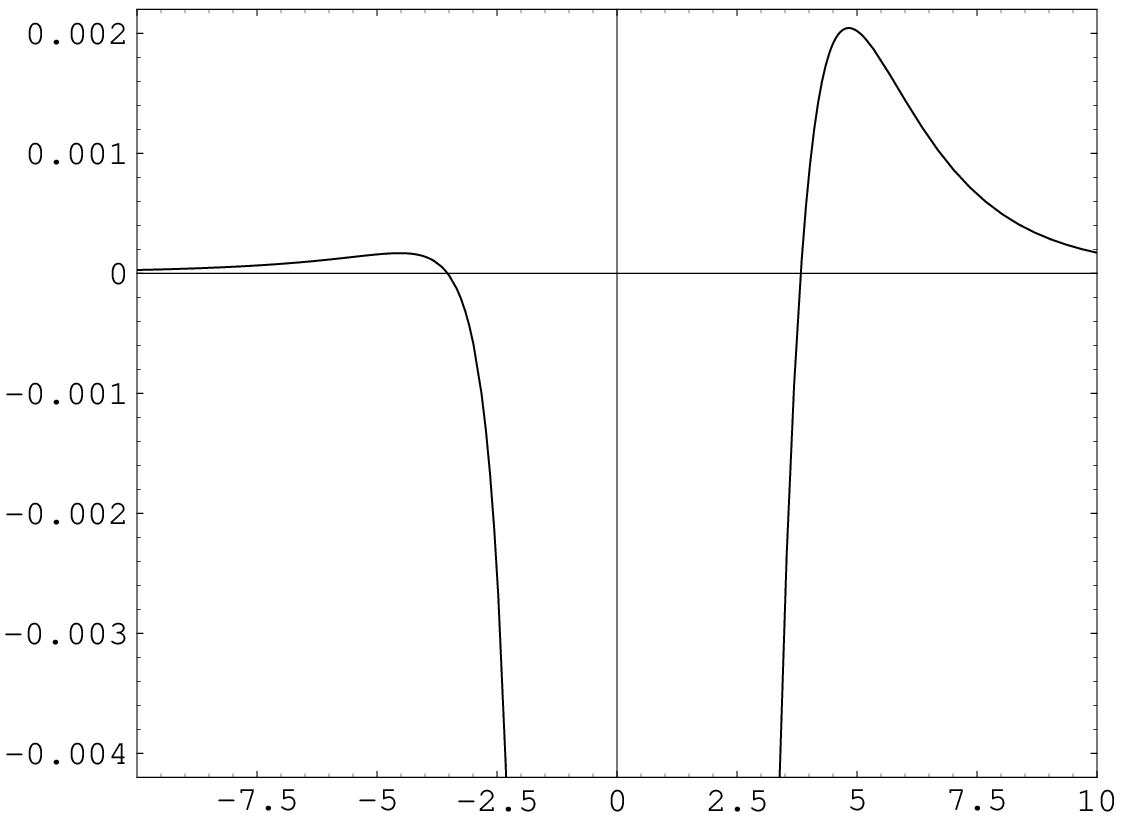,width=6cm,height=6cm}
\end{tabular}
\end{center}
\caption{Regularized Casimir energy density for a single plate
geometry
---formula {(\ref{eps1plate})}--- multiplied by $|\protect\beta _{1}|^{D+1}/(%
\protect\xi -\protect\xi _{c})$ as a function of $y$ for $D=1$
(left) and for $D=3$ (right).} \label{figEy}
\end{figure}

In Fig. \ref{figEy} we have plotted the dependence of $|\beta
_{1}|^{D+1}\varepsilon /(\xi -\xi _{c})$ on $y$ for the cases $D=1$ and $D=3$%
.
As we see, for a given $\beta _{1}$ the vacuum energy density has
a maximum for $x=x_{i}\equiv \beta _{1}y_{i}(D)$, $i=1,2$, where
$y=y_{i}$ correspond to the maxima in the figures and depend only
on the space dimension $D$. In Table \ref{tab1} we show the values
for these maxima and the corresponding values of the energy
density for $D=1,3,5,7$.

\bigskip

\begin{table}[tbp]
\begin{center}
\begin{tabular}{|c|c|c|c|c|}
\hline $D$ & $y_{1}$ & $y_{2}$ & $\left\langle T_{00}\right\rangle
_{SUB}(y_{1})$ & $\left\langle T_{00}\right\rangle _{SUB}(y_{2})$
\\ \hline
1 & 2.83 & -2.54 & 0.136 & 0.021 \\
3 & 4.83 & -4.53 & 2.04 10$^{-3}$ & 1.68 10$^{-4}$ \\
5 & 6.83 & -6.52 & 1.19 10$^{-5}$ & 6.68 10$^{-7}$ \\
7 & 8.83 & -8.51 & 4.44 10$^{-8}$ & 1.89 10$^{-9}$ \\ \hline
\end{tabular}
\end{center}
\caption{Maxima and corresponding values for the energy density
(multiplied by $|\beta _1|^{D+1}/(\xi -\xi _c)$) in the case of a
single plate.} \label{tab1}
\end{table}

\bigskip

\noindent When $\beta _{1}\rightarrow 0$, one has $x_{i}\rightarrow 0$ and $%
\left\langle T_{00}\right\rangle _{SUB}(x=x_{i})\sim \left| \beta
_{1}\right| ^{-D-1}\rightarrow +\infty $ and we obtain the
distribution corresponding to the Dirichlet boundary condition.

\subsection{Total vacuum energy} \label{sec:1pl2}

Integrating (\ref{eps1plate}) in $x$, we find for the energy per
unit surface area from $x=x_{1}$ to $x=\infty $,
\begin{eqnarray}
E_{x_{1}\leq x<\infty } &=&\int_{x_{1}}^{\infty }dx\left\langle
T_{00}\right\rangle _{SUB}=  \label{enx1inf} \\
&=&\frac{-D(\xi -\xi _{c})}{2^{D}\pi ^{D/2}\Gamma (D/2+1)\beta _{1}^{D}}%
\left[ \frac{\Gamma (D)}{y_{1}^{D}}-2e^{-y_{1}}\mbox{\rm Ei}%
(y_{1})+2\sum_{j=1}^{D-1}\frac{(j-1)!}{y_{1}^{j}}\right] ,\quad y_{1}=\frac{%
2x_{1}}{\beta _{1}}.  \nonumber
\end{eqnarray}
To investigate the total vacuum energy it is convenient to present
the subtracted vacuum EMT (\ref{reg1plate}) in another alternative
integral form. In the integral term on the right of
eq.(\ref{reg1plate}) we write the subintegrand in terms of
exponential functions and rotate the integration contour by angle
$\pi /2$ for the term with $e^{ik\left| x\right| }$ and by angle
$-\pi /2$ for the term with $e^{-ik\left| x\right| }$. This
procedure leads to the relation
\begin{eqnarray}
\int_{0}^{\infty }dkk^{D}\cos \left( 2kx+2\alpha _{1}(k)\right)  &=&-\sin
\frac{\pi D}{2}\,{\rm p.v.}\,\int_{0}^{\infty }dt\,t^{D}e^{-2t\left|
x\right| }\frac{\beta _{1}t+1}{\beta _{1}t-1}-  \label{introtform} \\
&&-2\pi \theta (\beta _{1})\frac{e^{-2\left| x\right| /\beta _{1}}}{\beta
_{1}^{D+1}}\cos \frac{\pi D}{2},  \nonumber
\end{eqnarray}
where the second summand on the right comes from the poles $\pm
i/\beta _{1}$ in the case $\beta _{1}>0$ and the symbol ${\rm
p.v.}$ means the principal value of the integral. Now comparing to
(\ref{reg1plate}) we see that this summand cancels the
contribution from the mode (\ref{imeig}). Substituting
(\ref{introtform}) into (\ref{reg1plate}) and using the gamma
function reflection formula
\begin{equation}
\sin \frac{\pi D}{2}\Gamma \left( -\frac{D}{2}\right) =-\frac{\pi
}{\Gamma (D/2+1)}  \label{rel4}
\end{equation}
for the subtracted v.e.v. one finds
\begin{equation}
\left\langle T_{ik}\right\rangle _{SUB}=\frac{D(\xi _{c}-\xi
)}{2^{D-1}\pi ^{D/2}\Gamma (D/2+1)}\,{\rm p.v.}\,\int_{0}^{\infty
}dt\,t^{D}e^{-2t\left| x\right| }\frac{\beta _{1}t+1}{\beta
_{1}t-1}\,\,{\rm diag}(1,0,-1,\cdots ,-1). \label{intform1pl1}
\end{equation}
Integrating the corresponding energy density for the total volume energy in
the region $0<x<\infty $ we receive
\begin{equation}
E^{{\rm (s)(vol)}}(\beta _{1})=\frac{D(\xi _{c}-\xi )}{2^{D}\pi ^{D/2}\Gamma
(D/2+1)}\,{\rm p.v.}\,\int_{0}^{\infty }dt\,t^{D-1}\,\frac{\beta _{1}t+1}{%
\beta _{1}t-1}\,.  \label{Esvol}
\end{equation}
Here and below the index (s) is for quantities describing the
single plate geometry.

In addition to volume energy (\ref{Esvol}) there is also surface
energy contribution to the total vacuum energy. The corresponding
energy density is defined by the relation (see, \cite{KCD})
\begin{equation}
T_{00}^{{\rm (surf)}}=-\frac{4\xi -1}{2}\delta (x;\partial
M)\varphi
\partial _{x}\varphi ,  \label{surfen1pl}
\end{equation}
where $\delta (x;\partial M)=\delta (x+0)$ is a ''one sided'' $\delta $ -
function. From this formula it follows that the surface term is zero for
Dirichlet or Neumann boundary condition (as the factors $\varphi $ or $%
\partial _{x}\varphi $ would then vanish) but yields a nonvanishing
contribution for Robin boundary conditions. The evaluation procedure for the
corresponding v.e.v. is similar to that given above for the volume part and
leads to the expression
\begin{eqnarray}
\langle 0|T_{00}^{{\rm (surf)}}|0\rangle  &=&(1/4-\xi )\frac{D\Gamma (-D/2)}{%
2^{D}\pi ^{D/2+1}}\delta (x+0)  \label{T00ssurf} \\
&&\times \left[ \int_{0}^{\infty }dkk^{D-1}\sin \left( 2kx+2\alpha
_{1}(k)\right) -2\pi \theta (\beta _{1})\frac{e^{-2x/\beta _{1}}}{\beta
_{1}^{D}}\cos \frac{\pi D}{2}\right] ,  \nonumber
\end{eqnarray}
where the second summand in the square brackets on the right comes from the
purely imaginary mode (\ref{imeig}). Transforming the integral term by the
way similar to (\ref{introtform}) and integrating the energy density over
the region $0\leq x<\infty $ for the total surface energy one finds
\begin{equation}
E^{{\rm (s)(surf)}}(\beta _{1})=\frac{D(\xi -1/4)}{2^{D}\pi ^{D/2}\Gamma
(D/2+1)}\,{\rm p.v.}\,\int_{0}^{\infty }dt\,t^{D-1}\,\frac{\beta _{1}t+1}{%
\beta _{1}t-1}\,.  \label{Essurf}
\end{equation}
As in the volumic part the contribution of mode (\ref{imeig}),
divergent for even values of $D$, is cancelled by the term coming
from the poles $\pm i/\beta _{1}$, $\beta _{1}>0$. The total
vacuum energy is the sum of the volumic and surface parts:
\begin{equation}
E^{{\rm (s)}}(\beta _{1})=-\frac{1}{2^{D+2}\pi ^{D/2}\Gamma (D/2+1)}\,{\rm %
p.v.}\,\int_{0}^{\infty }dt\,t^{D-1}\,\frac{\beta _{1}t+1}{\beta _{1}t-1}\,.
\label{Estot}
\end{equation}
The $\xi $ - dependence in the volumic and surface energies cancelled each
other and we have $\xi $ - independent total vacuum energy. The expressions (%
\ref{Esvol}), (\ref{Essurf}) and (\ref{Estot}) are divergent in given forms.
Due to the relations
\begin{equation}
E^{{\rm (s)(vol)}}=4D(\xi -\xi _{c})E^{{\rm (s)}},\quad E^{{\rm (s)(surf)}%
}=D(1-4\xi )E^{{\rm (s)}}  \label{Esreltot}
\end{equation}
it is sufficient to regularize the total vacuum energy. We will
return to this question in section \ref{sec:totcas}. By taking
into account that the regularized value of the integral
$\int_{0}^{\infty }dt\,t^{D-1}$ is equal to zero we conclude from
(\ref{Estot}) that the total regularized vacuum energy is zero for
Dirichlet and Neumann scalars (see also \cite{KCD}).

\section{Scalar Casimir effect with Robin boundary conditions on two
parallel plates} \label{sec:2pl}

In this section we will consider a scalar field with $\xi $
coupling and satisfying Robin boundary conditions (\ref{Rbcx0a}),
i.e.,
\begin{equation}
(1+\beta _{m}n^{\mu }\partial _{\mu })\varphi (t,{\bf x})=(1+\beta
_{m}(-1)^{m-1}\partial _{x})\varphi (t,{\bf x})=0,\quad x=a_{m},\quad m=1,2
\label{Robin2plate}
\end{equation}
on plane boundaries $x=a_{1}$ and $x=a_{2}$. The corresponding
eigenfunctions in the region between the plates can be presented in two
equivalent forms (corresponding to $m=1,2$)
\begin{equation}
\varphi _{{\bf k}}(t,{\bf x})=\beta e^{i{\bf k}_{\perp }{\bf x}-i\omega
t}\cos [k\left| x-a_{m}\right| +\alpha _{m}],  \label{eigenfunctions}
\end{equation}
where ${\bf k}=(k,{\bf k}_{\perp })$, $\omega =\sqrt{k_{\perp }^{2}+k^{2}}$,
and
\begin{equation}
\sin \alpha _{m}=\frac{1}{\sqrt{k^{2}\beta _{m}^{2}+1}},\quad \cos \alpha
_{m}=\frac{k\beta _{m}}{\sqrt{k^{2}\beta _{m}^{2}+1}}.  \label{alpha0}
\end{equation}
From the boundary conditions one obtains that the eigenmodes for $k$ are
solutions to the following equation
\begin{equation}
F(z)\equiv \left( 1-b_{1}b_{2}z^{2}\right) \sin z-(b_{2}+b_{1})z\cos
z=0,\quad z=ka,\quad b_{i}=\beta _{i}/a,\quad a=a_{2}-a_{1}.
\label{eigenmodes}
\end{equation}
In Appendix \ref{Append1} the values of the coefficients $b_{i}$
are specified for which this equation has purely imaginary zeros.
For these zeros the region of the allowed values for ${\bf
k}_{\perp }$ is restricted by the reality condition of $\omega $:
$k_{\perp }^{2}\geq -k^{2}$.

The coefficient $\beta $ is determined from the orthonormality condition
\begin{equation}
\int \varphi _{{\bf k}}\varphi _{{{\bf k^{\prime }}}}^{\ast }dV=\frac{1}{%
2\omega }\delta ({\bf k}_{\perp }-{{\bf k^{\prime }}_{\perp }})\delta
_{kk^{\prime }},  \label{orthonormality}
\end{equation}
where the integration goes over the region between the plates. Using the
form of the eigenfunctions one obtains
\begin{equation}
\beta ^{2}\omega a(2\pi )^{D-1}=\left| 1+\frac{1}{ka}\sin (ka)\cos
(ka+2\alpha _{m})\right| ^{-1}  \label{beta0}
\end{equation}
(on the class of the solutions $ka=z$ to (\ref{eigenmodes}) the expressions
on the right are the same for $m=1$ and $m=2$ ).

\subsection{Vacuum energy density and stresses} \label{sec:2pl1}

The v.e.v. for the EMT can be found by evaluating mode sum (\ref{vevEMT})
with the energy-momentum tensor (\ref{EMT}). Substituting the eigenfunctions
(\ref{eigenfunctions}) for the vacuum EMT components one finds
\begin{equation}
\langle 0|T_{ik}(x)|0\rangle =\frac{\delta _{ik}}{2}\sum_{z=\lambda
_{n},iy_{l}}\int \frac{d^{D-1}{\bf k}_{\perp }}{(2\pi )^{D-1}}\frac{%
B_{i}+A_{i}(z/a)\cos \left[ 2z\left| x-a_{m}\right| /a+2\alpha _{m}\right] }{%
\sqrt{k_{\perp }^{2}a^{2}+z^{2}}\left[ 1+\frac{1}{z}\sin z\cos (z+2\alpha
_{1})\right] },  \label{energydensity}
\end{equation}
where
\begin{equation}
B_{0}=k_{\perp }^{2}+z^{2}/a^{2},\quad B_{1}=z^{2}/a^{2},\quad B_{2}=\frac{%
k_{\perp }^{2}}{D-1},\quad i=2,...,D,  \label{Bi}
\end{equation}
and the coefficients $A_{i}(k)$ are defined as (\ref{Ai}) with $k=z/a$, $%
z=\lambda _{n}$ are positive solutions to equation (\ref{eigenmodes}), and $%
z=iy_{l}$, $y_{l}>0$ are the purely imaginary solutions in the upper
half-plane. For the latter case the ${\bf k}_{\perp }$-integration in
(\ref{energydensity}) extends over the region $k_{\perp }^{2}\geq y_{l}^{2}/a^{2}$%
.

Next, we apply to the sum over $n$ in eq.(\ref{energydensity}) the
formula derived in the appendix \ref{Append2}. Note that $\alpha
_{1}$ depends on $z=\lambda _{n},iy_l$ as well and
\begin{equation}
\cos \left( 2z\frac{\left| x-a_{m}\right| }{a}+2\alpha
_{m}\right) =\frac{z^{2}b_{m}^{2}-1}{z^{2}b_{m}^{2}+1}%
\cos \left( 2z\frac{x-a_{m}}{a}\right) -\frac{2zb_{m}}{%
z^{2}b_{m}^{2}+1}\sin \left( 2z\frac{\left| x-a_{m}\right|
}{a}\right) .  \label{cosalpha}
\end{equation}
In (\ref{energydensity}), we perform the integration over ${\bf
k}_{\perp }$ by using formula (\ref{int1}). Further, introducing a
new integration variable $y=k_{\perp }/k$ and evaluating the
corresponding integrals over $y$ using formulae (\ref{int2}) and
(\ref{int2im}), we find that the vacuum EMT has the form
(\ref{emtdiag}). Energy density, $\varepsilon $, pressures in
perpendicular, $p$, and parallel, $p_{\perp }$, to the plates
directions are determined by relations
\begin{equation}
q(x)=-\frac{\Gamma \left( -D/2\right) }{2^{D+1}\pi ^{D/2}a^{D+1}}%
{\sum_{z=\lambda _{n},\pm iy_{l}}}^{\prime }\frac{f_{m}^{(q)}(z,x)}{1+\frac{1}{z%
}\sin z\cos (z+2\alpha _{1})},\quad q=\varepsilon ,p,p_{\perp },
\label{qxplates}
\end{equation}
where the prime on the summation sign means that the contribution
of the terms corresponding to the purely imaginary zeros have to
be halved and the following notations are introduced
\begin{equation}
\begin{array}{lll}
f_{m}^{(\varepsilon )}(z,x) & = & \displaystyle z^{D}\left\{ 1+\frac{4D(\xi
-\xi _{c})}{z^{2}b_{m}^{2}+1}\left[ \left( z^{2}b_{m}^{2}-1\right) \cos
\left( 2z\frac{x-a_{m}}{a}\right) -2zb_{m}\sin \left( 2z\frac{\left|
x-a_{m}\right| }{a}\right) \right] \right\} , \\
f_{m}^{(p)}(z,x) & = & Dz^{D}, \\
f_{m}^{(p_{\perp })}(z,x) & = & -f_{m}^{(\varepsilon )}(z,x).
\end{array}
\label{fnzx}
\end{equation}
It follows from here that $p_{\perp }=-\varepsilon $, and, for the
conformally coupled scalar (i.e. $\xi =\xi _{c}$), the components of the
vacuum EMT are uniform between the plates. Similarly to the cases of
Dirichlet and Neumann boundary conditions, these properties can be also
directly obtained by using symmetry arguments. The field equation and
boundary conditions are invariant with respect to the Lorentz boosts in
directions parallel to the plates. It follows from here that the
corresponding (transverse to $x$ axis) part of the vacuum EMT is
proportional to the metric tensor, and hence
\begin{equation}
\varepsilon =\langle 0|T_{0}^{0}(x)|0\rangle =\langle
0|T_{2}^{2}(x)|0\rangle =...=\langle 0|T_{D}^{D}(x)|0\rangle =-p_{\perp }.
\label{epspe}
\end{equation}
For the conformally invariant field, from the zero trace condition one finds
$p=-\langle 0|T_{1}^{1}(x)|0\rangle =D\varepsilon $. By the symmetry of the
problem, the quantities $q$ depend only on $x$ coordinate. From the
continuity equation for the EMT it follows that $p^{\prime }(x)=0$, and
therefore the vacuum EMT is constant. Unlike the case of Dirichlet and
Neumann boundary conditions, for the conformally coupled scalar the
functional dependence on the plates separation cannot be determined by
purely dimensional arguments, because we have three parameters with length
dimension, $a,\beta _{1},\beta _{2}$. To obtain the dependence on these
parameters we need an explicit calculation.

The summation in (\ref{qxplates}) can be done by using  formula (\ref
{sumformula}) taking $f(z)=f_{m}^{(q)}(z,x)$. As a result one obtains
\begin{eqnarray}
{\sum_{z=\lambda _{n},\pm iy_{l}}}^{\prime }\frac{\pi
f_{m}^{(q)}(z,x)}{1+\frac{\sin z}{z}\cos (z+2\alpha _{1})} & = & %
\displaystyle\int_{0}^{\infty }f_{m}^{(q)}(z,x)dz-2\sin \left( \frac{\pi D}{2%
}\right) {\rm p.v.}\int_{0}^{\infty }\frac{F_{m}^{(q)}(t,x)dt}{\frac{%
(b_{1}t-1)(b_{2}t-1)}{(b_{1}t+1)(b_{2}t+1)}e^{2t}-1} \nonumber \\
&  & +8\pi \cos \left( \frac{\pi D}{2}\right) \frac{D(\xi -\xi
_{c})}{b_{m}^{D+1}}\theta \left( b_{m}\right) \exp \left(
-2\frac{\left| x-a_{m}\right| }{ab_{m}}\right) , \label{sum2plate}
\end{eqnarray}
where the following notations have been introduced
\begin{equation}
\begin{array}{lll}
F_{m}^{(\varepsilon )}(z,x) & = & \displaystyle z^{D}\left\{ 1+\frac{4D(\xi
-\xi _{c})}{z^{2}b_{m}^{2}-1}\left[ \left( z^{2}b_{m}^{2}+1\right) \cosh
\left( 2z\frac{x-a_{m}}{a}\right) -2zb_{m}\sinh \left( 2z\frac{\left|
x-a_{m}\right| }{a}\right) \right] \right\}  \\
F_{m}^{(p)}(z,x) & = & Dz^{D}, \\
F_{m}^{(p_{\perp })}(z,x) & = & -F_{m}^{(\varepsilon )}(z,x).
\end{array}
\label{Fmzx}
\end{equation}
Note that on the right of (\ref{sum2plate}) we have included the term (\ref
{frompoles}) coming from the poles $\pm i/b_{m}$ for $b_{m}>0$ .

By comparing to the formula (\ref{reg1plate}) it can be easily seen that the
contribution of the first and last summands on the right of formula (\ref
{sum2plate}) corresponds to the vacuum EMT components for the geometry of a
single plate placed at $x=a_{m}$. This can be also seen by taking the limit $%
(-1)^{m^{\prime }}a_{m^{\prime }}\rightarrow \infty $, $m^{\prime }\neq m$, $%
m^{\prime }=1,2$, when the second integral on the right of eq. (\ref
{sum2plate}), multiplied by $a^{-D-1}$, tends to zero. The regularization
for the single plate case was carried out in previous section. Hence, using
formula (\ref{sum2plate}) for the regularized v.e.v. of the EMT for the case
of two plate geometry from (\ref{qxplates}) one obtains
\begin{equation}
{\rm reg\,}q(x)={\rm reg\,}q_{m}^{{\rm (s)}}(x)-\frac{2^{-D}\pi ^{-D/2}}{%
a^{D+1}\Gamma \left( D/2+1\right) }{\rm p.v.}\int_{0}^{\infty }\frac{%
F_{m}^{(q)}(t,x)dt}{\frac{(b_{1}t-1)(b_{2}t-1)}{(b_{1}t+1)(b_{2}t+1)}e^{2t}-1%
}.  \label{qx2}
\end{equation}
Here we have used the gamma function reflection formula
(\ref{rel4}). In (\ref{qx2}) the term ${\rm reg\,}q_{m}^{{\rm
(s)}}(x)$ is the regularized v.e.v. for the case of a single plate
placed at $x=a_{m}$. This geometry was investigated in the
previous section and the corresponding regularized quantities are
given by relations (\ref{eps1plate}) and (\ref
{regp1plate}) with the replacement $x\rightarrow x-a_{m}$. As follows from (%
\ref{Fmzx}) and (\ref{qx2}) the vacuum perpendicular pressure, ${\rm reg\,}p$%
, is uniform in the region between the plates:
\begin{equation}
{\rm reg}\,\,p=D\,\varepsilon _{c}^{(1)},\quad \varepsilon _{c}^{(1)}\equiv -%
\frac{2^{-D}\pi ^{-D/2}}{a^{D+1}\Gamma (D/2+1)}{\rm p.v.}\int_{0}^{\infty }%
\frac{t^{D}dt}{\frac{(b_{1}t-1)(b_{2}t-1)}{(b_{1}t+1)(b_{2}t+1)}e^{2t}-1}.
\label{regpe}
\end{equation}
The Casimir forces per unit area on the plates are equal to this
vacuum pressure and are attractive for negative values of ${\rm
reg}\,\, p$, and repulsive for positive values. In the limit
$|\beta _1 |/a\ll 1$ using the value of the integral
\begin{equation}
\int_{0}^{\infty }\frac{t^{D}dt}{e^{2t}-1}=\frac{\zeta _{R}(D+1)}{2^{D+1}}%
\Gamma (D+1)  \label{inttDdt}
\end{equation}
($\zeta _{R}$ meaning the Riemann zeta function) from
(\ref{regpe}) one has
\begin{equation}\label{regplimD}
  {\rm reg}\,\,p=({\rm reg}\,\,p)_{{\rm Dirichlet}}
  \left[ 1+\frac{\beta _1+\beta _2}{a}(D+1)+\cdots \right] ,
\end{equation}
where
\begin{equation}\label{regpDir}
({\rm reg}\,\,p)_{{\rm Dirichlet}}=-\frac{D\zeta _R(D+1)}{(4\pi
)^{(D+1)/2}a^{D+1}}\Gamma \left( \frac{D+1}{2}\right)
\end{equation}
is the corresponding vacuum pressure for the case of the Dirichlet
boundary condition \cite{Ambjorn}. The latter corresponds to the
attractive vacuum force for any value of $D$. Similarly, in the
limit $|\beta _1 |/a\gg 1$ one derives
\begin{equation}\label{regplimN}
{\rm reg}\,\,p=({\rm reg}\,\,p)_{{\rm Neumann}}
  \left[ 1+4a\left( \frac{1}{\beta _1}+\frac{1}{\beta _2}\right)
  \frac{\zeta _R(D-1)}{D\zeta _R(D+1)}+\cdots \right] ,
\end{equation}
with $({\rm reg}\,\,p)_{{\rm Neumann}}=({\rm reg}\,\,p)_{{\rm
Dirichlet}}$ being the vacuum pressure for the Neumann scalar. For
given Robin coefficients $\beta _i$ formula (\ref{regplimD}) gives
the asymptotic behaviour of the vacuum forces for large values
$a$, and eq. (\ref{regplimN}) gives the asymptotic for small
distances. In both these limits the vacuum forces are attractive.

\begin{figure}[htbp]
\begin{center}
\begin{tabular}{c}
\psfig{figure=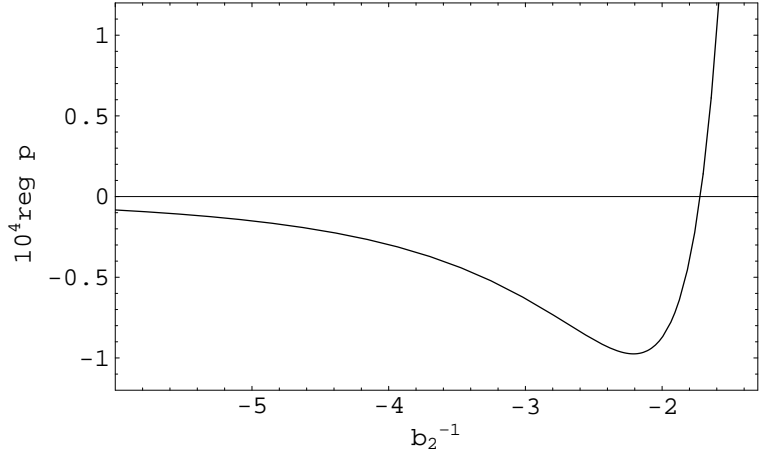,width=10cm,height=6cm}
\end{tabular}
\end{center}
\caption{Casimir force per unit surface  multiplied by $10^4\beta _2^4$%
, i.e. $10^4\beta _2^4 \protect {\rm reg}\,\, p$ as a function of
$b_2^{-1}=a/\beta _2$, for $D=3$, $\beta _1=0$}
\label{fig2plforce}
\end{figure}

Let us consider in detail the special case $\beta _1=0$ (Dirichlet
condition on the plate $x=a_1$). Now in the limit of large
distances, $|\beta _2 |/a\ll 1$, the vacuum forces are dominated
by term (\ref{regpDir}) and, hence, are attractive. At small
distances, $|\beta _2 |/a\gg 1$, from (\ref{regpe}) with $\beta
_1=0$ one has
\begin{equation}\label{regpbet10}
{\rm reg}\,\,p=-\left( 1-2^{-D}\right) ({\rm reg}\,\,p)_{{\rm
Dirichlet}}\left( 1+{\cal O}(a/|\beta _2|)\right) ,
\end{equation}
which corresponds to the repulsive force. It follows from here
that at some intermediate value of $a$ the Casimir force vanishes.
As a result we have an example when the vacuum forces stabilize
the distance between the plates. This feature is seen from
Fig.\ref{fig2plforce}, where the Casimir force is plotted versus
$a/\beta _2$ in the case $D=3$ and $\beta _1=0$.

In (\ref{qx2}) taking the limit $(-1)^{m}a_{m}\rightarrow \infty $ one
obtains integral form (\ref{intform1pl1}) for the v.e.v. in the case of a
single plate. By using this formula  the corresponding quantities (\ref{qx2}%
) for two plate geometry can be presented as
\begin{equation}
{\rm reg\,}q(x)={\rm reg\,}q^{{\rm (s)}}(x;a_{1})+{\rm reg\,}q^{{\rm (s)}%
}(x;a_{2})+\Delta q(x),  \label{qa1a2int}
\end{equation}
where ''interference '' term has the form
\begin{equation}
\Delta q(x)=-\frac{2^{-D}\pi ^{-D/2}}{a^{D+1}\Gamma \left( D/2+1\right) }%
{\rm p.v.}\int_{0}^{\infty }\frac{\Delta F^{(q)}(t,x)dt}{\frac{%
(b_{1}t-1)(b_{2}t-1)}{(b_{1}t+1)(b_{2}t+1)}e^{2t}-1}  \label{iterference}
\end{equation}
with notation
\begin{equation}
\Delta F^{(\varepsilon )}(t,x)=t^{D}\left[ 1+2D(\xi -\xi _{c})\sum_{m=1}^{2}%
\frac{b_{m}t+1}{b_{m}t-1}\exp \left( -2t\frac{\left|
x-a_{m}\right| }{a}\right) \right] . \label{DeltaF}
\end{equation}
In (\ref{qa1a2int}) the quantities $\Delta q(x)$ are finite for all values $%
a_{1}\leq x\leq a_{2}$ and the surface divergences at $x=a_{m}$
are contained in the summands ${\rm reg\,}q^{{\rm (s)}}(x;a_{m})$.
Note that the
quantities (\ref{iterference}) are symmetric under the replacement $%
b_{1}\leftrightarrows b_{2}$.

For the conformally coupled scalar field ${\rm reg\,}q_{m}^{{\rm (s)}}(x)=0$%
, and for the vacuum EMT components between the plates from (\ref{qx2}) one
obtains
\begin{equation}
\begin{array}{lll}
{\rm reg}\,\,\varepsilon  & = & {\rm reg}\,\,p/D=\varepsilon _{c}^{(1)} \\
{\rm reg\,}p_{\perp } & = & -{\rm reg}\,\,\varepsilon ,
\end{array}
\label{regconf}
\end{equation}
where $\varepsilon _{c}^{(1)}$ is determined from (\ref{regpe}). Hence $%
\varepsilon _{c}^{(1)}$ is the vacuum energy density between the
plates for the conformally coupled scalar field.

In the case $b_{1}=-b_{2}$ using the value of integral
(\ref{inttDdt}) and the duplication formula
\begin{equation}
\Gamma (2z)=\frac{2^{2z-1}}{\sqrt{\pi }}\Gamma (z)\Gamma \left( z+\frac{1}{2}%
\right) ,  \label{Gammadup}
\end{equation}
eq.(\ref{qx2}) turns into
\begin{equation}
{\rm reg}\,\varepsilon =-\frac{\zeta _{R}(D+1)}{(4\pi )^{(D+1)/2}a^{D+1}}%
\Gamma \left( \frac{D+1}{2}\right) .  \label{endensDir}
\end{equation}
This formula coincides with that derived in \cite{Ambjorn} for the Dirichlet
and Neumann cases. Notice that, whenever $b_{1}=-b_{2}\equiv b$, the
particular value of $b$ does not matter, as one might have been observed at
the beginning, from the form of the $F$ function.

\subsection{Total volume energy} \label{sec:2pl2}

From\bigskip\ relation (\ref{qa1a2int}) for the total volume energy per unit
surface one has
\begin{equation}
E^{{\rm (vol)}}=\int_{a_{1}}^{a_{2}}dx\,{\rm reg}\,\varepsilon =E^{{\rm (s)}%
}(a_{1}<x<a_{2};\beta _{1})+E^{{\rm (s)}}(a_{1}<x<a_{2};\beta _{2})+\Delta E,
\label{Evol}
\end{equation}
where $E^{{\rm (s)}}(a_{1}<x<a_{2};\beta _{m})$ is the vacuum energy in the
region $a_{1}<x<a_{2}$ due to a single plate at $x=a_{m}$ with Robin
boundary condition (\ref{Robin2plate}). The ''interference'' term $\Delta E$
$=\int_{a_{1}}^{a_{2}}dx\,\Delta \varepsilon (x)$ can be presented in the
form
\begin{equation}
\Delta E=-\frac{2^{-D}\pi ^{-D/2}}{a^{D}\Gamma \left( D/2+1\right) }{\rm p.v.%
}\int_{0}^{\infty }\frac{t^{D}dt}{\frac{(b_{1}t-1)(b_{2}t-1)}{%
(b_{1}t+1)(b_{2}t+1)}e^{2t}-1}\left[ 1+D(\xi -\xi _{c})\frac{1-e^{-2t}}{t}%
\sum_{m=1}^{2}\frac{b_{m}t+1}{b_{m}t-1}\right] .  \label{DeltaE}
\end{equation}
The energy $E^{{\rm (s)}}(a_{1}<x<a_{2};\beta _{1})$ can be obtained
subtracting from the total volume energy of a single plate (given by (\ref
{Esvol})) the vacuum energy in the region $a_{2}\leq x<\infty $. The latter
can be easily derived integrating the 00-component of eq. (\ref{intform1pl1}%
). As a result one finds
\begin{equation}
E^{{\rm (s)}}(a_{1}<x<a_{2};\beta _{1})=E^{{\rm (s)(vol)}}(\beta _{1})+\frac{%
D(\xi -\xi _{c})}{2^{D}\pi ^{D/2}\Gamma (D/2+1)a^{D}}\,{\rm p.v.}%
\,\int_{0}^{\infty }dt\,t^{D-1}e^{-2t}\frac{\beta _{1}t+1}{\beta _{1}t-1}.
\label{Esa1a2}
\end{equation}
Taking into account this and similar formula for $E^{{\rm (s)}%
}(a_{1}<x<a_{2};\beta _{2})$ from (\ref{Evol}) and (\ref{DeltaE}) we receive
\begin{equation}
E^{{\rm (vol)}}=\sum_{m=1}^{2}E^{{\rm (s)(vol)}}(\beta _{m})+a\left(
\varepsilon _{c}^{(1)}+4D(\xi -\xi _{c})\varepsilon _{c}^{(2)}\right) ,
\label{Evol1}
\end{equation}
where $\varepsilon _{c}^{(1)}$ is defined as in (\ref{regpe}), and we have
introduced notation
\begin{equation}
\varepsilon _{c}^{(2)}={\frac{b_{1}+b_{2}}{2^{D}\pi ^{D/2}a^{D+1}\Gamma
\left( 1+{\frac{D}{2}}\right) }}\,{\rm p.v.}\int_{0}^{\infty }dt\,{\frac{%
t^{D}(1-b_{1}b_{2}t^{2})}{(1-b_{1}t)^{2}(1-b_{2}t)^{2}%
\,e^{2t}-(1-b_{1}^{2}t^{2})(1-b_{2}^{2}t^{2})}.}  \label{epsc2}
\end{equation}
For Dirichlet ($b_{1}=b_{2}=0$) and Neumann ($b_{1}=b_{2}=\infty $) scalars
this term vanishes. Note that the volume energy (\ref{Evol1}) is symmetric
under the replacement $b_{1}\leftrightarrows b_{2}$.

\section{Total Casimir energy} \label{sec:totcas}

Up to now, we have obtained local energy densities from expectation values
of the energy-momentum tensor. Here, we will take the alternative approach
of calculating the integrated Casimir energy per unit volume ---$\varepsilon
_{c}$--- from the eigenvalue sum, namely,
\begin{equation}
\varepsilon _{c}={\frac{1}{2a}}\sum_{n}\int {\frac{d^{D-1}{\bf k}_{\perp }}{%
(2\pi )^{D-1}}}\,\sqrt{k_{\perp }^{2}+k_{n}^{2}},  \label{defEc}
\end{equation}
where, $k_{n}\equiv {z}_{n}/a$, being $\{z_{n}=\lambda _{n},iy_{n}\}$ the
set of the zeros of the $F(z)$ function defined by eq.(\ref{eigenmodes}). In
the case of the purely imaginary zeros the ${\bf k}_{\perp }$-integration
ranges over the region $k_{\perp }^{2}\geq y_{n}^{2}/a^{2}$.

\subsection{Zeta function regularization} \label{sec:zetfunc}

As it stands, the r.h.s. of (\ref{defEc}) clearly diverges, but we shall
evaluate by zeta function regularization, i.e., turning (\ref{defEc}) into
the function
\begin{equation}
\varepsilon _{c}(\mu ;s)=\,{\frac{\mu }{2a}}\sum_{n}\int {\frac{d^{D-1}{\bf k%
}_{\perp }}{(2\pi )^{D-1}}}\,\left( \frac{k_{\perp }^{2}+k_{n}^{2}}{\mu ^{2}}%
\right) ^{-s/2}.  \label{defEcs}
\end{equation}
and adopting the prescription that the regularized value of $\varepsilon
_{c} $ will be $\varepsilon _{c}(\mu ;s=-1)$ (some of the pionnering works
in this sort of technique are listed as ref.\cite{ztrg}). Note that we have
introduced the arbitrary mass scale $\mu $ in order to keep dimensionless
the quantity raised to the power of $-s/2$. After integrating over ${\bf k}%
_{\perp }$, eq. (\ref{defEcs}) may be written as follows
\begin{equation}
\varepsilon _{c}(\mu ;s)=\,{\frac{\mu ^{s+1}}{2^{D}\,\pi ^{(D-1)/2}\,a^{D-s}}%
}{\frac{\Gamma \left( {\frac{\sigma }{2}}\right) }{\Gamma \left( \frac{s}{2}%
\right) }}\left[ \zeta _{\Lambda }(\sigma )-\frac{\cos (\pi
D/2)}{\sin (\pi s/2)}\sum_{l}y_{l}^{-\sigma }\right] ,\quad \sigma
=s+1-D , \label{EcZL}
\end{equation}
where
\begin{equation}
\zeta _{\Lambda }(\sigma )=\sum_{n=1}^{\infty }\lambda _{n}^{-\sigma }
\label{defZL}
\end{equation}
is the zeta function for the real zeros of $F(z)$.

First, we look at the asymptotic form of the $\lambda _{n}$'s in order to
find the convergence boundary of $\zeta _{\Lambda }(\sigma )$. If $%
b_{1}b_{2}\neq 0$, then $F(z)\sim b_{1}b_{2}z^{2}\sin z$ as $|z|\gg 1$, and $%
\displaystyle\zeta _{\Lambda }(\sigma )\sim \sum_{n\geq 1}(\pi n)^{-\sigma
}=\pi ^{-\sigma }\zeta _{R}(\sigma )$ ---$\zeta _{R}$ denoting the Riemann
zeta function---, which has a pole at $\sigma =1$. When $b_{1}=0$ and $%
b_{2}\neq 0$ (or $b_{1}\neq 0$ and $b_{2}=0$), one has $F(z)\sim
-b_{2,1}z\cos z$ for $|z|\gg 1$. Thus, $\displaystyle\zeta
_{\Lambda }(\sigma )\sim \sum_{n}(\pi (n+1/2))^{-\sigma }=\pi
^{-\sigma }(2^{\sigma }-1)\zeta
_{R}(\sigma )$ which has a pole at $\sigma =1$, too. Finally, if $%
b_{1}=b_{2}=0$, $F(z)=\sin z$ and, therefore, $\displaystyle\zeta _{\Lambda
}(\sigma )=\pi ^{-\sigma }\zeta _{R}(\sigma )$ exactly. In view of this, we
realize that $\zeta _{\Lambda }(\sigma )$ has its rightmost pole at $\sigma
=1$, regardless of the values of $b_{1}$, $b_{2}$.

This way, we see that eq. (\ref{defZL}) is, initially, just valid for the
domain Re $\sigma >1$. Nevertheless, in order to obtain the regularized
Casimir energy, one has to find the analytic continuation of (\ref{defZL})
to $\sigma =-D$, which corresponds to $s=-1$ (here, $D=1,2,3,\dots $). This
task will be done by analytic extension of an adequate contour integration
in the complex plane. An immediate consequence of the Cauchy formula for the
residues of a complex function is the expression
\begin{equation}
\begin{array}{lllc}
\zeta _{\Lambda }(\sigma ) & = & \displaystyle{\frac{1}{2\pi i}}%
\int_{C}\,dz\,z^{-\sigma }\,{\frac{d}{dz}}\ln F(z), &
\mbox{ for Re $\sigma
>1$},
\end{array}
\label{crZL}
\end{equation}
where $C$ is a closed circuit enclosing all the zeros of $F(z)$. In this
case, we assume that $C$ is made of a large semicircle ---with radius
tending to infinity--- centered at the origin and placed to its right, plus
a straight part overlapping the imaginary axis, which avoids the origin, the
possible purely imaginary zeros $\pm iy_{l}$, $y_{l}>0$, and the points $\pm
i/b_{1}$, $\pm i/b_{2}$ by small semicircles whose radii tend to zero.
However, the contribution from the small semicircle around the origin will
vanish when we manage to shift the initial $\sigma $-domain to the left of
its initial position and $\sigma $ becomes negative enough to reach $-D$.
Bearing this in mind, we may already neglect this part.

The asymptotic behaviour of the $F$ function on the upper and lower
half-planes motivates the factorization
\begin{equation}
F(z)=F_{1}(z)\,F_{2}(z),
\end{equation}
where
\begin{equation}
\begin{array}{l}
F_{1}(z)=
\begin{array}{llll}
F_{1}^{\pm }(z) & \equiv & \displaystyle-{\frac{i}{2}}\,(1\pm ib_{1}z)(1\pm
ib_{2}z)\,e^{\mp iz} & \mbox{for Im$(z) { > \atop < } 0$,}
\end{array}
\\
F_{2}(z)=
\begin{array}{llll}
F_{2}^{\pm }(z) & \equiv & \displaystyle1-{\frac{(1\mp ib_{1}z)(1\mp ib_{2}z)%
}{(1\pm ib_{1}z)(1\pm ib_{2}z)}}\,e^{\pm 2iz} &
\mbox{for Im$(z) { >
\atop < } 0$.}
\end{array}
\end{array}
\end{equation}
On this basis, the original integration is decomposed as follows
\begin{equation}
\begin{array}{lll}
\displaystyle\int_{C}dz\,z^{-\sigma }\frac{d}{dz}\,\mbox{ln}\ F(z) & = & %
\displaystyle\int_{C}dz\,z^{-\sigma }\frac{d}{dz}\,\mbox{ln}\
F_{1}(z)+\int_{C}dz\,z^{-\sigma }\frac{d}{dz}\,\mbox{ln}\ F_{2}(z) \\
\displaystyle\int_{C}dz\,z^{-\sigma }\frac{d}{dz}\,\mbox{ln}\ F_{2}(z) & = & %
\displaystyle\int_{C^{+}}dz\,z^{-\sigma }\frac{d}{dz}\,\mbox{ln}\
F_{2}^{+}(z)+\int_{C^{-}}dz\,z^{-\sigma }\frac{d}{dz}\,\mbox{ln}\
F_{2}^{-}(z),
\end{array}
\label{decomp}
\end{equation}
with $C^{+}$ and $C^{-}$ denoting the upper and lower halves of the
integration circuit, which have Im$(z)>0$ and Im$(z)<0$, respectively. By
virtue of our definitions of $F_{2}^{\pm }$,
\begin{equation}
\begin{array}{llll}
\mbox{ln}\ F_{2}^{\pm }(z) & = & {\cal O}\left( e^{{\pm }2iz}\right) & %
\mbox{for Im$(z) { > \atop < } 0$.}
\end{array}
\end{equation}
Thanks to these properties, the integrals involving $\mbox{ln}\ F_{2}^{\pm }$
will vanish on the large circular parts as their common radius ---say $R$---
tends to infinity, i.e., separating $\displaystyle\int_{C^{\pm
}}=\int_{A^{\pm }}+\int_{V^{\pm }},$ where ${A^{\pm }}$ stand for {\it Arcs}
and ${V^{\pm }}$ for {\it Verticals}, we have that $\displaystyle%
\int_{A^{\pm }}dz\,z^{-\sigma }\frac{d}{dz}\,\mbox{ln}\ F_{2}^{\pm
}(z)\longrightarrow 0,$ as $R\longrightarrow \infty $ and, in consequence,
\begin{equation}
\begin{array}{c}
\displaystyle{\frac{1}{2\pi i}}\left[ \int_{C^{+}}dz\,z^{-\sigma }\frac{d}{dz%
}\,\mbox{ln}\ F_{2}^{+}(z)+\int_{C^{-}}dz\,z^{-\sigma }\frac{d}{dz}\,%
\mbox{ln}\ F_{2}^{-}(z)\right] \\
\displaystyle={\frac{1}{2\pi i}}\left[ \int_{V^{+}}dz\,z^{-\sigma }\frac{d}{%
dz}\,\mbox{ln}\ F_{2}^{+}(z)+\int_{V^{-}}dz\,z^{-\sigma }\frac{d}{dz}\,%
\mbox{ln}\ F_{2}^{-}(z)\right] \\
\displaystyle=\zeta _{\Lambda }^{(0)}(\sigma )+\Delta \zeta _{\Lambda
}(\sigma ). \\
\end{array}
\label{intCplCmin}
\end{equation}
Here we use the notations
\begin{equation}
\begin{array}{rcl}
\displaystyle\zeta _{\Lambda }^{(0)}(\sigma ) & \equiv & \displaystyle{\frac{%
1}{\pi }}\sin \left( \frac{\pi \sigma }{2}\right) \,\mbox{p.v.}%
\,\int_{0}^{\infty }dt\,t^{-\sigma }\frac{d}{dt}\,\mbox{ln}\ \left[ 1-%
\displaystyle{\frac{(1+b_{1}t)(1+b_{2}t)}{(1-b_{1}t)(1-b_{2}t)}}\,e^{-2t}%
\right] \\
\displaystyle\Delta \zeta _{\Lambda }(\sigma ) & \equiv & \displaystyle%
\left\{ \sum_{m=1}^{2}\theta (\,b_{m})b_{m}^{\sigma
}-\sum_{l}y_{l}^{-\sigma }\right\} \cos \left( \frac{\pi \sigma
}{2}\right) .
\end{array}
\label{decomp1}
\end{equation}
 Observe that the result denoted by
$\zeta _{\Lambda }^{(0)}(\sigma )$ has been obtained after parametrizing the
$V^{\pm }$ segments in the way: $z=e^{i\pi /2}t$ with $t$ from $\infty $ to
0, for $V^{+}$, and $z=e^{-i\pi /2}t$ with $t$ from 0 to $\infty $, for $%
V^{-}$. The $\Delta \zeta _{\Lambda }(\sigma )$ term comes from the
integrals over semicircles avoiding the purely imaginary zeros and poles of $%
F_{2}^{\pm }(z)$.

The remaining piece in (\ref{decomp}) can be evaluated similarly:
\begin{eqnarray}
\frac{1}{2\pi i}\int_{C}dz\,z^{-\sigma }\frac{d}{dz}\,\mbox{ln}\ F_{1}(z) &=&%
\frac{1}{\pi }\sin \frac{\pi \sigma }{2}\,\mbox{p.v.}\int_{0}^{\infty
}dz\,z^{-\sigma }\left( 1+\sum_{m=1}^{2}\frac{1}{z-1/b_{m}}\right) -
 \nonumber \\
&&-\sum_{m=1}^{2}\theta (b_{m})b_{m}^{\sigma }\cos \frac{\pi
\sigma }{2}, \label{decomp2}
\end{eqnarray}
where the second summand on the right comes from the poles $\pm i/b_{m}$ for
$b_{m}>0$. Now we see that this term cancels the corressponding one in the
expression for $\Delta \zeta _{\Lambda }(\sigma )$. Putting together all
these contributions ---eqs.(\ref{crZL}), (\ref{intCplCmin}), (\ref{decomp1}), (%
\ref{decomp2})---, we arrive at
\begin{equation}
\zeta _{\Lambda }(\sigma )=\zeta _{\Lambda }^{(0)}(\sigma )-\cos \frac{\pi
\sigma }{2}\sum_{l}y_{l}^{-\sigma }+\frac{1}{\pi }\sin \frac{\pi \sigma }{2}%
\,\mbox{p.v.}\int_{0}^{\infty }dz\,z^{-\sigma }\left( 1+\sum_{m=1}^{2}\frac{1%
}{z-1/b_{m}}\right) .  \label{zetlam}
\end{equation}
From eqs. (\ref{EcZL}) and (\ref{zetlam}) one derives
\begin{eqnarray}
\varepsilon _{c}(\mu ;s) &=&\frac{\mu ^{s+1}a^{-1}}{2^{D}\pi
^{(D-1)/2}\Gamma (s/2)\Gamma (1-\sigma /2)}\left\{{\rm cotan\,}\frac{\pi s%
}{2}\sum_{l}\left( \frac{y_{l}}{a}\right) ^{-\sigma
}+\mbox{p.v.}\int_{0}^{\infty }dt\,t^{-\sigma
}\times  \right. \nonumber   \\
&\times & \left.\left[ a+\sum_{m=1}^{2}\frac{1}{t-1/\beta _{m}}+\frac{d}{dt}\,%
\mbox{ln}\ \left| 1-\displaystyle{\frac{(1+\beta _{1}t)(1+\beta _{2}t)}{%
(1-\beta _{1}t)(1-\beta _{2}t)}}\,e^{-2at}\right| \right] \right\}
.
 \label{epscmus4}
\end{eqnarray}
It can be seen that the contribution to the vacuum energy
corresponding to the first term in the square brackets under the
integral is the vacuum energy density for the Minkowski vacuum
without boundaries and its regularized value is equal to zero.
Hence, we will omit this term in the
following consideration. The contribution of the complex zeros vanishes at $%
s=-1$ due to the factor ${\rm cotan}(\pi s/2)$. The contribution
of the
last, logarithmic, term is finite at $s=-1$ and vanishes in the limit $%
a\rightarrow \infty $. It follows from here that the term
\begin{equation}
E^{{\rm (s)}}(\mu ;s)=a\varepsilon _{c}^{{\rm (s)}}(\mu ;s)=\frac{\mu
^{s+1}2^{-D}\pi ^{(1-D)/2}}{\Gamma \left( \frac{s}{2}\right) \Gamma \left(
\frac{D+1-s}{2}\right) }\mbox{p.v.}\int_{0}^{\infty }dt\,\frac{t^{D-1-s}}{%
t-1/\beta _{m}}  \label{tot1pl3}
\end{equation}
corresponds to the total vacuum energy in the region $a_{m}\leq
x<\infty $ for the case of a single plate with Robin coefficient
$\beta _{m}$. By
taking into account that the regularized value of the integral $%
\int_{0}^{\infty }dt\,t^{D-1}$ is zero we see that $E^{{\rm
(s)}}(\mu ;-1)$ coincides with the total vacuum energy $E^{{\rm
(s)}}$ derived in section \ref{sec:2pl} by summing the
corresponding volume and surface energies. The integral in formula
(\ref{tot1pl3}) can be evaluated by using the standard formulae
(see, for instance, \cite{Prudnikov}):
\begin{equation}
E^{{\rm (s)}}(\mu ;s)=\frac{2^{-D}\pi ^{(3-D)/2}}{\left| \beta _{m}\right|
^{D}\Gamma \left( \frac{s}{2}\right) \Gamma \left( \frac{D+1-s}{2}\right) }%
\frac{(\mu \left| \beta _{m}\right| )^{s+1}}{\sin (D-s)\pi }\left\{
\begin{array}{cc}
1, & \beta _{m}<0 \\
-\cos (D-s)\pi , & \beta _{m}>0
\end{array}
\right. .  \label{tot1pl5}
\end{equation}
As we see the total vacuum energy has a pole at $s=-1$. Laurent-expanding
where necessary, for the pole structure one derives
\begin{equation}
E^{{\rm (s)}}(\mu ;s)=-\frac{2^{-D-1}\pi ^{-D/2}}{\beta _{m}^{D}\Gamma
(D/2+1)}\left\{ \frac{1}{s+1}+\ln (\mu \left| \beta _{m}\right| )+\frac{1}{2}%
\psi \left( \frac{D}{2}+1\right) -\frac{1}{2}\psi \left( -\frac{1}{2}\right)
+{\cal O}(s+1)\right\} ,  \label{tot1pl7}
\end{equation}
where $\psi (s)=d\ln \Gamma (s)/ds$ is the diagamma function. We
will assume that the pole term is absorbed by the corresponding
part in the bare action \cite{KCD}. The way in which the pole is
removed is not unique. The renormalization scheme of ref.
\cite{Wipf} corresponds to the minimal subtraction and is
equivalent to simply removing the pole term from eq. (\ref
{tot1pl7}). In this way for the finite part of the total energy of
a single plate we receive
\begin{equation}
E^{{\rm (s)}}(\beta _{m})=-\frac{2\ln (\mu \left| \beta _{m}\right| )+\psi
\left( D/2+1\right) -\psi \left( -1/2\right) }{2^{D+2}\pi ^{D/2}\beta
_{m}^{D}\Gamma (D/2+1)}.  \label{tot1plren}
\end{equation}
This energy depends on the arbitrary scale $\mu $, as usually
happens with this type of regularization. The discussion for the
role of this scale can be found in \cite{Wipf}. The corresponding
volume and surface energies are obtained by using relations
(\ref{Esreltot}). Note that the term containing the normalization
scale is independent on the distance between the plates and does
not contribute to the vacuum forces.

Now from eq.(\ref{epscmus4}),  for $\sigma =-(D-1-s)$, for the total vacuum
energy in the region $a_{1}\leq x\leq a_{2}$ one derives
\begin{equation}
E=\sum_{m=1}^{2}E^{{\rm (s)}}(\beta _{m})+a\varepsilon _{c}^{(0)},
\label{resEc11}
\end{equation}
where
\begin{equation}
\varepsilon _{c}^{(0)}=\frac{-1}{2^{D+1}\pi ^{D/2}\Gamma \left( D/2+1\right)
a^{D+1}}\,{\rm p.v.}\,\int_{0}^{\infty }dt\,t^{D}\frac{d}{dt}\ln \left[ 1-%
\frac{(1+b_{1}t)(1+b_{2}t)}{(1-b_{1}t)(1-b_{2}t)}\,e^{-2t}\right]
. \label{resEc1}
\end{equation}
Further, performing a differentiation in the integral in eq.(\ref{resEc1}),
one obtains the following decomposition
\begin{equation}
\begin{array}{lll}
\displaystyle\varepsilon _{c}^{(0)} & = & \varepsilon _{c}^{(1)}+\varepsilon
_{c}^{(2)}
\end{array}
,  \label{eps1eps2}
\end{equation}
where $\varepsilon _{c}^{(1)}$ and $\varepsilon _{c}^{(2)}$ are defined in (%
\ref{regpe}) and (\ref{epsc2}). Now, comparing this with eqs.(\ref{regconf}%
), (\ref{regpe}), we recognize $\varepsilon _{c}^{(1)}$ as the volumic part
of the integrated energy per unit-volume of a conformally coupled field,
i.e., the part coming just from the volume between the plates, already
calculated in the previous section. Since $\varepsilon _{c}^{(0)}$ accounts
for the ''interference'' part of \ the {\it total} integrated energy per
unit-volume, the difference, $\varepsilon _{c}^{(0)}-\varepsilon
_{c}^{(1)}=\varepsilon _{c}^{(2)}$, has to be identified as the contribution
from the surfaces of the plates.

\subsection{Identification in terms surface density} \label{sec:surfen}

Next, we may consider the implications of this fact in terms of the
densities found in the previous section. From relation (\ref{energydensity}%
), for the integrated Casimir energy per unit volume in the region {\it %
between} the plates, one obtains
\begin{equation}
\begin{array}{lll}
\varepsilon _{c}^{{\rm (vol)}} & = & \displaystyle\frac{1}{a}%
\int_{a_{1}}^{a_{2}}dx\,\langle 0\left| T_{00}\right| 0\rangle  \\
& = & \displaystyle\frac{1}{2a}\sum_{z=\lambda _{n},iy_{l}}\int \frac{d^{D-1}%
{\bf k}_{\perp }}{(2\pi )^{D-1}}\left[ \sqrt{k_{\perp }^{2}+z^{2}/a^{2}}+%
\frac{(4\xi -1)z}{a^{2}\sqrt{k_{\perp }^{2}+z^{2}/a^{2}}}\frac{\sin z\cos
(z+2\alpha _{1})}{1+\frac{1}{z}\sin z\cos (z+2\alpha _{1})}\right] ,
\end{array}
\label{inten1}
\end{equation}
where for the purely imaginary zeros the ${\bf k}_{\perp }$ - integration
goes over the region $k_{\perp }^{2}\geq y_{l}^{2}/a^{2}$. As we see, this
result differs from the total Casimir energy per unit volume (\ref{defEc}).
We have argued that the reason for this difference should be the existence
of an additional surface energy contribution to the volume energy (\ref
{inten1}), located on the boundaries $x=a_{m}$, $m=1,2$. The corresponding
energy density is defined by relation (\ref{surfen1pl}), where now
\begin{equation}
\quad \delta (x;\partial M)=\delta (x-a_{1}+0)-\delta (x-a_{2}-0).
\label{surfen1}
\end{equation}
The corresponding v.e.v. can be evaluated by the standard method explained
in the previous section. This leads to the formula
\begin{equation}
\begin{array}{lll}
\langle 0\left| T_{00}^{{\rm (surf)}}\right| 0\rangle  & = & \displaystyle%
\frac{4\xi -1}{2}\delta (x;\partial M)\left( \partial _{x}\langle 0\left|
\varphi (x)\varphi (x^{\prime })\right| 0\rangle \right) \mid _{x^{\prime
}=x}= \\
& = & \displaystyle\frac{4\xi -1}{4a^{2}}\delta (x;\partial
M)\sum_{z=\lambda _{n},iy_{l}}\int \frac{d^{D-1}{\bf k}_{\perp }}{(2\pi
)^{D-1}}\frac{\lambda _{n}}{\sqrt{k_{\perp }^{2}+z^{2}/a^{2}}}\frac{\sin
\left[ 2z\frac{x-a_{1}}{a}+2\alpha _{1}\right] }{1+\frac{1}{z}\sin z\cos
(z+2\alpha _{1})},
\end{array}
\label{surfendens}
\end{equation}
which provides the energy density on the plates themselves. Then, for the
integrated surface energy per unit area one obtains
\begin{equation}
\begin{array}{rcl}
\varepsilon _{c}^{{\rm (surf)}} & = & \displaystyle\frac{1}{a}%
\int_{a_{1}}^{a_{2}}dx\,\langle 0\left| T_{00}^{{\rm (surf)}}\right|
0\rangle  \\
& = & \displaystyle-\frac{4\xi -1}{2a^{3}}\sum_{z=\lambda _{n},iy_{l}}\int
\frac{d^{D-1}{\bf k}_{\perp }}{(2\pi )^{D-1}}\frac{z}{\sqrt{k_{\perp
}^{2}+z^{2}/a^{2}}}\frac{\sin z\cos (z+2\alpha _{1})}{1+\frac{1}{z}\sin
z\cos (z+2\alpha _{1})}.
\end{array}
\label{surfen}
\end{equation}
Adding up (\ref{inten1}) and (\ref{surfen}), one re-obtains the standard
result (\ref{defEc}). Thus, we have just checked that $\varepsilon
_{c}=\varepsilon _{c}^{{\rm (vol)}}+\varepsilon _{c}^{{\rm (surf)}}$. After
the standard integration over transverse momentum eq. (\ref{surfen}) may be
written as follows
\begin{equation}
\varepsilon _{c}^{{\rm (surf)}}=\frac{4\xi -1}{a^{D+1}}\frac{D\Gamma (-D/2)}{%
2^{D+1}\pi ^{D/2}}\left[ {\sum_{z=\lambda _{n},\pm
iy_{l}}}^{\prime }z^{D}-{\sum_{z=\lambda _{n},\pm iy_{l}}}^{\prime
}\frac{z^{D}}{1+\frac{1}{z}\sin z\cos (z+2\alpha _{1})}\right] .
\label{surfen2}
\end{equation}
The regularized value for the first sum in the square brackets
have been found in the previous subsection (see (\ref{decomp1})
and (\ref{zetlam})). The second sum might be evaluated using the
Abel-Plana summation formula. Applying formula (\ref{sumformula})
to this sum and omitting the divergent contribution from the
integral term (this corresponds to the calculation of
(\ref{surfendens}) taking in this formula $\langle \varphi
(x)\varphi (x^{\prime })\rangle _{SUB}=\langle 0\left| \varphi
(x)\varphi (x^{\prime })\right| 0\rangle -\langle 0_{M}\left|
\varphi (x)\varphi (x^{\prime })\right| 0_{M}\rangle $ instead of
$\langle 0\left| \varphi (x)\varphi (x^{\prime })\right| 0\rangle
$) for the surface energy per unit area one obtains
\begin{equation}
E^{{\rm (surf)}}=a\varepsilon _{c}^{{\rm (surf)}}=\sum _{m=1}^{2}E^{{\rm (s)(surf)}%
}(\beta _m)-aD(4\xi -1)\varepsilon _{c}^{(2)}  \label{surfen3}
\end{equation}
with $\varepsilon _{c}^{(2)}$ defined in (\ref{epsc2}). Now using (\ref
{eps1eps2}) and (\ref{surfen3}) we can find the volume part of the vacuum
energy as
\begin{equation}
E^{{\rm (vol)}}=E^{{\rm (tot)}}-E^{{\rm (surf)}}=\sum
_{m=1}^{2}E^{{\rm (s)(vol)}}(\beta _m)+a\left( \varepsilon
_{c}^{(1)}+4D(\xi -\xi _{c})\varepsilon _{c}^{(2)}\right) ,
\label{volen3}
\end{equation}
which coincides with the previous result (\ref{Evol1}) obtained by
integrating energy density. Hence, we have shown that the local
and global approaches lead to the same expression for the volume
energy in the case of the scalar field with general conformal
coupling $\xi $. Note that, as it follows from (\ref{surfen3}),
the quantity $\varepsilon _{c}^{(2)}$ is the additional (to the
single plate) surface energy per unit volume in the case of the
conformally coupled scalar field.

\subsection{Numerical examples} \label{sec:numex}

\begin{figure}[htbp]
\begin{center}
\begin{tabular}{c}
\psfig{figure=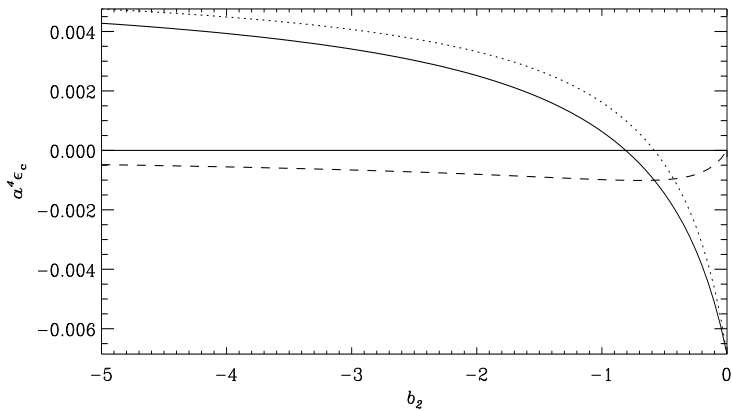,width=10cm,height=6cm}
\end{tabular}
\end{center}
\caption{Total integrated Casimir energy per unit-volume multiplied by $a^4$%
, i.e. $a^4 \protect\varepsilon_c$, for $D=3$, $b_1=0$, and $-5
\le b_2 \le 0 $. Separately plotted are the total value, in solid
line, the volumic contribution, in dotted line, and the
superficial part, in dashed line. Note the minimum of this surface
contribution at $b_2\simeq -0.70$, and the zeros
of the total value, at $b_2\simeq -0.81$, and of the volume part, at $%
b_2\simeq -0.58$. } \label{figb10}
\end{figure}

Formulas (\ref{resEc1}) and (\ref{eps1eps2}) with (\ref{regpe}), (\ref{epsc2}%
) are suitable for numerical evaluation (actually, we have used an
alternative form of $\varepsilon _{c}^{(0)}$ found by partial
integration of the integral in eq.(\ref{resEc1})). Making use of
these expressions, we may, e.g., keep the value of $b_{1}$ fixed
and study the variation with $b_{2}$ of the total ---$\varepsilon
_{c}^{(0)}$---, volumic ---$\varepsilon _{c}^{(1)}$---, and
superficial ---$\varepsilon _{c}^{(2)}$--- integrated Casimir
energies per unit-volume in the conformal case. An example for
$D=3$ is given in Fig. \ref{figb10}, where we have set $b_{1}=0$
while $b_{2}$ changes. The curve in solid line depicts
$a^{4}\varepsilon _{c}^{(0)}$ and the one in dotted line
$a^{4}\varepsilon _{c}^{(1)}$, being the surface contribution
$a^{4}\varepsilon _{c}^{(2)}$ the difference between them, which
has been plotted in dashed line. Note that, at $b_{2}=0$, one
recovers the known result $a^{4}\,\varepsilon =-{\frac{1}{16\pi
^{2}}}\,\Gamma (2)\,\zeta _{R}(4)=-{\frac{\pi ^{2}}{1440}}\simeq
-0.0069$ for the total and volumic parts, while the superficial
contribution is zero. The volume part is higher than the total
result, meaning that the surface contribution is
always negative. In fact, the magnitude of the latter tends to zero when $%
b_{2}\rightarrow -\infty $, as had to be expected, because this
limit corresponds to Neumann boundary conditions. Thus, between
$b_{2}=0$ and that asymptotic regime it must have at least one
minimum, and we actually observe one at $b_{2}\simeq -0.70$. To be
remarked is the zero of $\varepsilon _{c}^{(0)}$ at $b_{2}\simeq
-0.81$. Below this value, the total Casimir effect is repulsive
while, for larger values of $b_{2}$ until $b_{2}=0$, is
attractive.

\begin{figure}[htbp]
\begin{center}
\begin{tabular}{ccc}
\psfig{figure=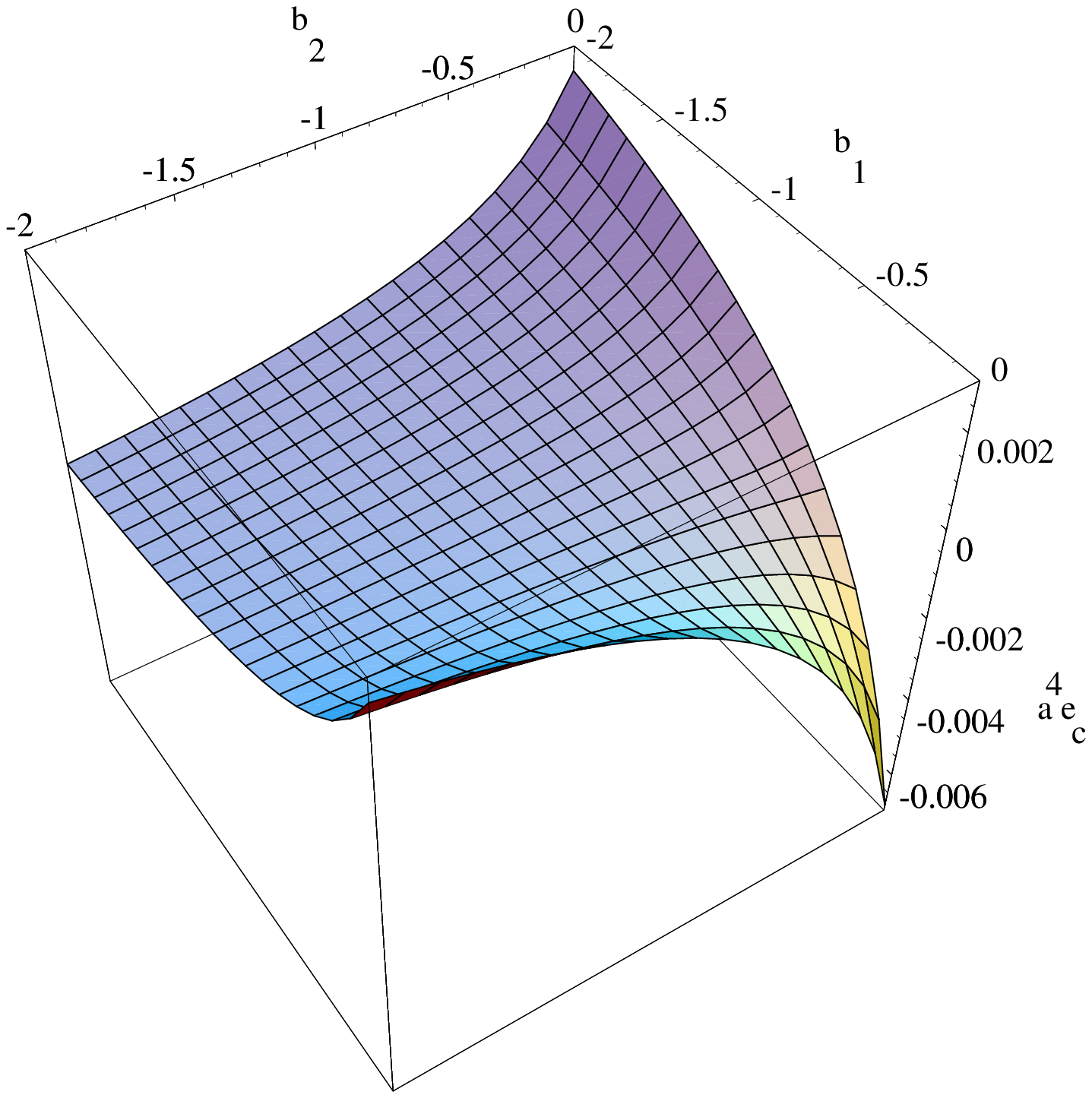,width=7cm,height=8cm} & \hspace*{1cm} & %
\psfig{figure=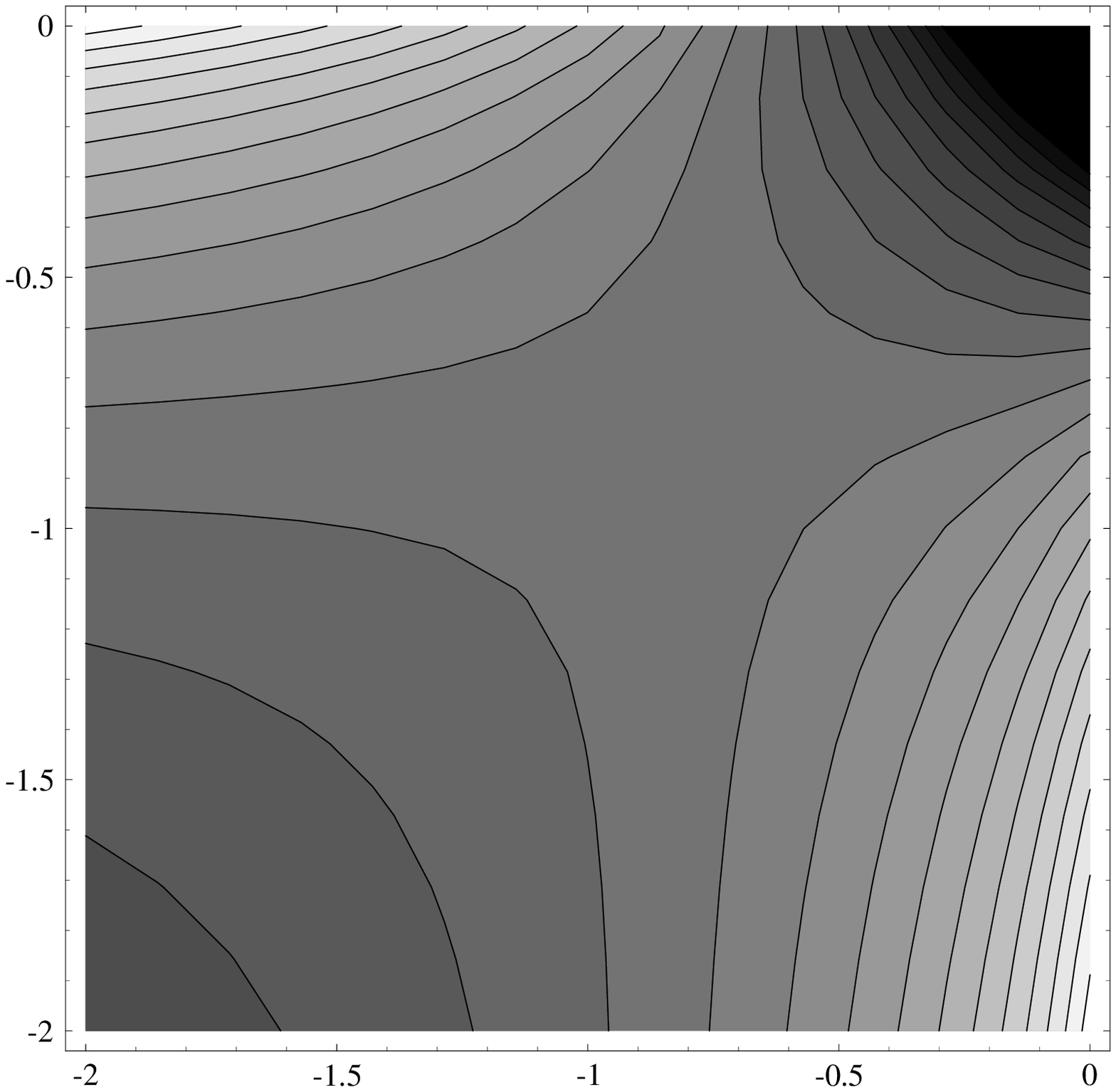,width=5cm,height=5cm}
\end{tabular}
\end{center}
\caption{ Left: Total integrated Casimir energy per unit-volume
times $a^4$, i.e. $a^4 \,\protect\varepsilon_c$, corresponding to
$D=3$, $-2 \le b_1 \le 0 $, $-2 \le b_2 \le 0$ (in this region,
there are no imaginary eigenfrequencies and the p.v. prescription
for the integrals is unnecessary). Right: contour representation
of the same plot.} \label{figb1b2}
\end{figure}

\begin{figure}[htbp]
\begin{center}
\begin{tabular}{ccc}
\psfig{figure=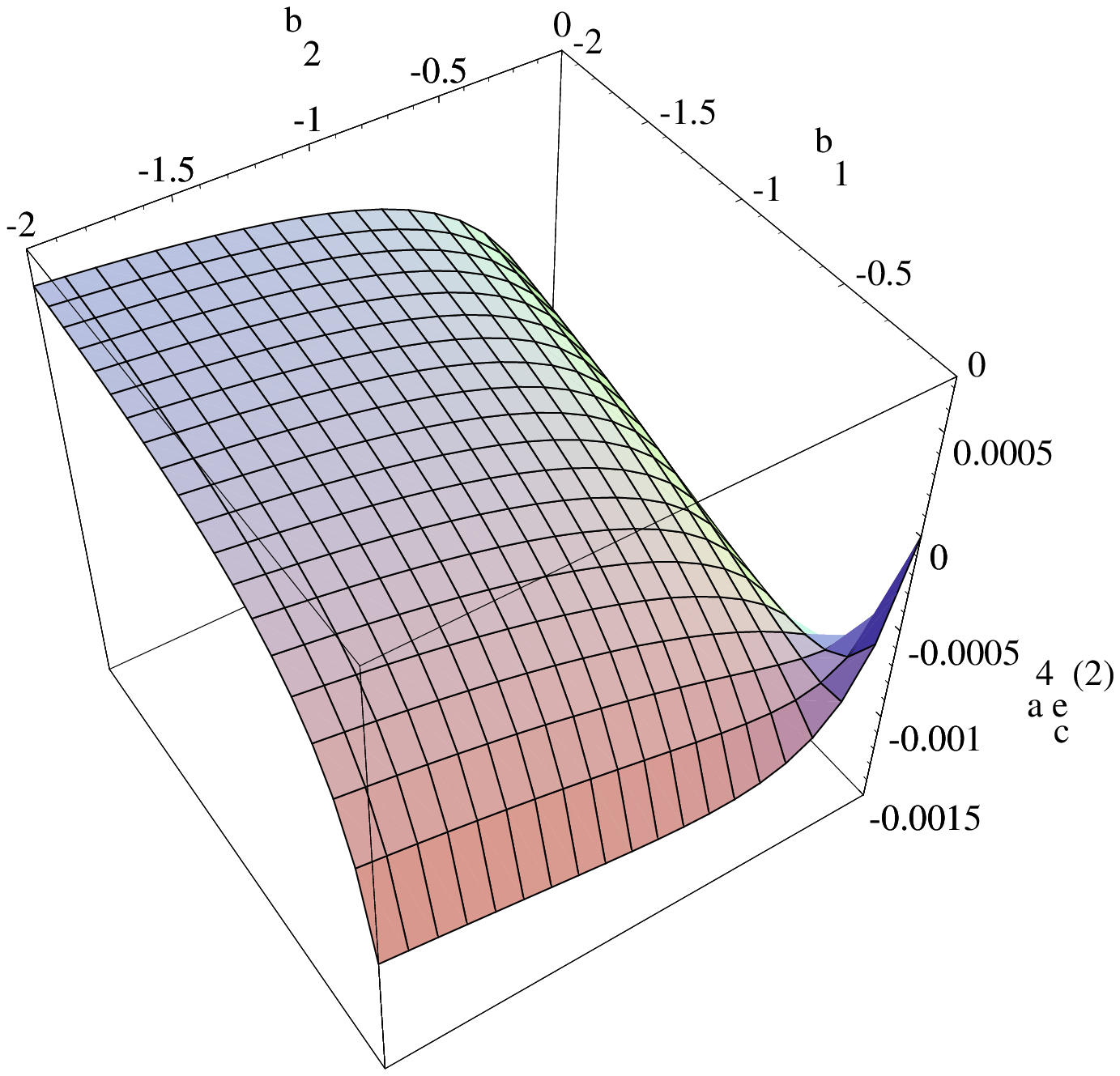,width=7cm,height=8cm} & \hspace*{1cm} & %
\psfig{figure=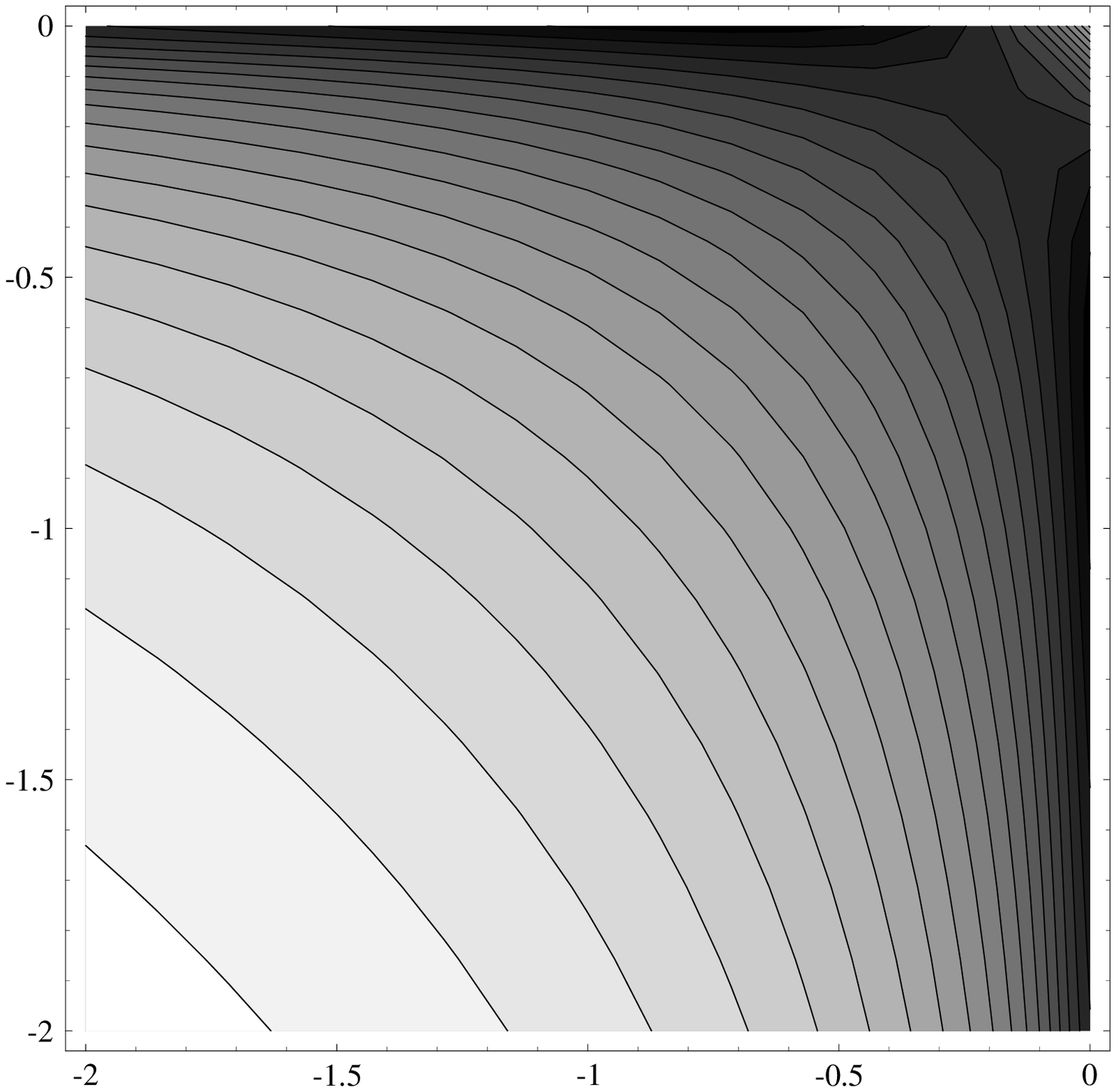,width=5cm,height=5cm}
\end{tabular}
\end{center}
\caption{Left: surface contribution $a^4
\,\protect\varepsilon_c^{(2)}$, for
$D=3$, $-2 \le b_1 \le 0$, $-2 \le b_2 \le 0$. For $b_1=0$, the minimum at $%
b_2\simeq -0.70$ is visible. Right: associated contour plot.}
\label{fige2}
\end{figure}

It is also possible to consider the Casimir energy as a function
of $b_1, b_2 $ simultaneously. The plot shown in Fig.
\ref{figb1b2} illustrates the changing form of the total energy
$\varepsilon_c\equiv\varepsilon_c^{(0)}$ on a given region of the
$(b_1,b_2)$-plane (for $D=3$, too). An analogous description of
$\varepsilon_c^{(2)}$ is provided by Fig. \ref{fige2}.


\section{Ending comments} \label{sec:endcom}

In the present work we have dealt with a calculation of the
Casimir energy when one sets Robin boundary conditions on one
single plate or a pair of parallel plates. Its evaluation has been
based on a variant of the generalized Abel-Plana summation formula
in ref.\cite{Sahrev}, adapted to these situations, and derived in
the appendix B. This method turns out to be adequate for finding
vacuum expectation values of the energy-momentum tensor, i.e.,
local densities. From a slightly different viewpoint, zeta
function regularization has been applied to the summation of
eigenfrequencies, which directly gives the integrated energy per
unit-volume.

When just one plate is considered, the only present parameter is
the relative coefficient between the non-derivative and derivative
terms in the boundary condition ($\beta_1$). The local density is
given by formula (\ref {eps1plate}), which vanishes for the
conformal value of the curvature coupling. Otherwise, this formula
depicts the local dependence of this density (exemplified by
Fig.\ref{figEy} for $D=1$ and $D=3$), which is singular on the
plate itself. Note that the requirement of conformal invariance
has the power of suppressing the presence of divergent parts, just
as happened ---for a different system--- in ref.\cite{ER}.

If there are two parallel plates, the relevant parameters are
three: the (rescaled) relative coefficients between the
non-derivative and derivative parts at each boundary ($b_{1}$ and
$b_{2}$), and the separation length between them ($a$). Then, the
additional (to a single plate) total integrated Casimir energy per
unit-volume is given by formula (\ref{resEc1}), and its
decomposition into purely-volume and purely-surface parts by eqs.
(\ref{Evol1}) and (\ref {surfen3}), in terms of the quantities
$\varepsilon ^{(1)}$ and $\varepsilon ^{(2)}$ defined by eqs.
(\ref{regpe}) and (\ref{epsc2}). If the coupling is conformal
($\xi =\xi _{c}$), $\varepsilon ^{(1)}$ and $\varepsilon ^{(2)}$
themselves coincide with the volume and surface contributions,
respectively, and, in any case, the decomposition (\ref{eps1eps2})
holds. The surface contribution, coming from the plates
themselves, would be absent for Dirichlet or Neumann boundary
conditions.

To be remarked is the fact that, at least in some situations free
of imaginary eigenfrequencies, there are parameter choices which
give a vanishing Casimir energy. As illustrated by Fig.
\ref{figb10}, one may vary the value of the $b_2$ parameter so as
to reverse the sign of the effect. At the same time, we have seen
that there is another $b_2$-value for which the
surface contribution has a minimum. Examples of simultaneous variations of $%
b_1$ and $b_2$ are shown in Figs. \ref{figb1b2} and \ref{fige2}.

An interesting feature of the Casimir effect with Robin boundary
conditions is that there is a region in the space of parameters
defining the boundary conditions in which the vacuum forces are
repulsive for small distances and attractive for large distances.
This leads to the possibility for the stabilization of the plates
separation by using the Casimir effect.

\section*{Acknowledgements}

The work of AAS was supported by the Armenian National Science and
Education Fund (ANSEF) Grant No. PS14-00.

\appendix

\section{Appendix: Complex zeros} \label{Append1}

\bigskip First of all we will show that the real and possible purely
imaginary zeros (see below) of $F(z)$ are simple. To see this, we
note that on the class of solutions to (\ref{eigenmodes}), the
corresponding derivative can be presented in the form
\begin{equation}
F^{\prime }(z)=\left[ 1+\frac{1}{z}\sin z\cos (z+2\alpha
_{1})\right] \left[ \left( 1-b_{1}b_{2}z^{2}\right) \cos
z+(b_{2}+b_{1})z\sin z\right] . \label{Fderiv}
\end{equation}
Using the integral relation
\begin{equation}
1+\frac{1}{z}\sin z\cos (z+2\alpha _{1})=2\int_{0}^{1}dx\,\cos
^{2}(zx+\alpha _{1})  \label{intrelap1}
\end{equation}
we conclude from here that $F^{\prime }(z)\neq 0$ if $z$, $z\neq
0$ is a zero of $F(z)$, and hence these zeros are simple.

Purely imaginary zeros of $F(z)$ may exist. This sort of solution
has to do with the presence of imaginary parts in the
eigenfrequencies. They can be detected as the real zeros of the
denominator in the last integral of eq.(\ref {sumformula}). It is
convenient to write the corresponding equation in the form
\begin{equation}
\tanh t=\frac{(b_{1}+b_{2})t}{1+b_{1}b_{2}t^{2}}.
\label{equation}
\end{equation}
After studying the nature of this equation in terms of $b_{1}$ and
$b_{2}$, one finds out that:

\noindent {\bf 1)} Equation (\ref{equation}) has no positive real
zeros for
\begin{equation}
\left\{ b_{1}+b_{2}\geq 1,\ b_{1}b_{2}\leq 0\right\} \cup \left\{
b_{1,2}\leq 0\right\} .  \label{nozeros}
\end{equation}

\noindent {\bf 2)} Equation (\ref{equation}) has a single positive
real zero for
\begin{equation}
\left\{ 0<b_{1}+b_{2}<1,b_{1}b_{2}\leq 0\right\} \cup \left\{
b_{1}+b_{2}\geq 1,b_{1,2}>0\right\} \cup \left\{
b_{1}+b_{2}<0,b_{1}b_{2}<0\right\} .  \label{1zero}
\end{equation}

\noindent {\bf 3)} Equation (\ref{equation}) has two positive real
zeros for
\[
b_{1}+b_{2}<1,\ b_{1,2}>0.
\]

The parameter values in Fig. \ref{figb10}, namely $b_{1}=0$ and
$-5\leq b_{2}\leq 0$, fall into case 1), when there are no real
positive zeros. When $F(z)$ has complex zeros (situations 2 or 3),
their extra contribution to the mode sum is given by formula
(\ref{frompoles}) in appendix B.

\section{Appendix: Summation formula} \label{Append2}

The vacuum expectation values for the physical quantities in the region
between plates will contain the sums over zeros of the function $F(z)$
defined by (\ref{eigenmodes}). To obtain the summation formula over these
zeros we will use the generalized Abel-Plana formula (GAPF) \cite{Sahrev}.
In this formula as a function $g(z)$ let us choose
\begin{equation}
g(z)=-i\left[ \left( 1-b_{1}b_{2}z^{2}\right) \cos z+(b_{2}+b_{1})z\sin z%
\right] \frac{f(z)}{F(z)}.  \label{gezet}
\end{equation}
For the sum and difference in the GAPF one has
\begin{equation}
g(z)\pm f(z)=i(b_{1}z\pm i)(b_{2}z\pm i)e^{\pm iz}\frac{f(z)}{F(z)}.
\label{gepmf}
\end{equation}
Let us denote by $\lambda _{n}$, $n=1,2,\ldots $ the zeros of the function $%
F(z)$ in the right half-plane, arranged in ascending order, and by $\pm
iy_{l}$, $y_{l}>0$ the possible purely imaginary zeros of this function. It
can be easily seen that
\begin{equation}
{\rm Res}_{z=\lambda _{n}}\, g(z)=\frac{-if(z)}{1+\frac{1}{z}\sin
z\cos (z+2\alpha _{1})}  \label{resge}
\end{equation}
(as it follows from (\ref{intrelap1}) the denominator on the right of this
formula is always positive). First, we will consider the case of function $%
f(z)$ analytic for ${\rm Re}z\geq 0$. Now substituting in GAPF (formulas
(2.10)-(2.11) in \cite{Sahrev}) (\ref{gezet}), (\ref{gepmf}), taking the
limit $a\rightarrow 0$ (here $a$ is the parameter on the right of GAPF and
the poles $\pm iy_{l}$ are excluded by small semicircles with radius $\rho $
on the right half plane, $\rho \rightarrow 0$), and using (\ref{resge}) one
obtains the following summation formula
\begin{eqnarray}
{\sum _{z=\lambda _{n},\pm iy_{l}}}^{\prime }\frac{\pi f(z)}{1+\frac{1}{%
z}\sin z\cos (z+2\alpha _{1})}&=&-\frac{\pi }{2}\frac{f(0)}{1-b_{2}-b_{1}}%
+\int_{0}^{\infty }f(z)dz+ \nonumber \\
 &+& i\,{\rm
p.v.}\int_{0}^{\infty }\frac{\left[
f(it)-f(-it)\right] dt}{\frac{(b_{1}t-1)(b_{2}t-1)}{(b_{1}t+1)(b_{2}t+1)}%
e^{2t}-1}dt, \label{sumformula}
\end{eqnarray}
where the prime on the summation sign means that the contribution
of terms corresponding to the purely imaginary zeros have to be
halved. This contribution comes from the integrals taken around
semicircles enclosing these zeros. Note that the denominator on
the left can be also written in the form
\begin{equation}
1+\frac{1}{z}\sin z\cos (z+2\alpha _{1})=1-\frac{%
(b_{1}+b_{2})(1+b_{1}b_{2}z^{2})}{(1+b_{1}^{2}z^{2})(1+b_{2}^{2}z^{2})}%
,\quad z=\lambda _{n},\pm iy_{l}.  \label{denom}
\end{equation}
In (\ref{sumformula}) we have assumed that $b_{1}+b_{2}\neq 1$. In the case $%
b_{1}+b_{2}=1$ to ensure the convergence at origin in the second integral on
the right of formula (\ref{sumformula}) we need to have $f(z)\sim
f_{0}z^{\alpha }$, $\alpha \geq 2$, $z\rightarrow 0$. Now the first summand
on the right of this formula should be replaced by
\begin{equation}
-\frac{\pi f_{0}\delta _{\alpha 2}}{2(b_{1}^{2}-b_{1}+1/3)}.
\label{sumformterm}
\end{equation}

Formula (\ref{sumformula}) is valid for functions $f(z)$ satisfyng the
condition
\begin{equation}
\left| f(z)\right| <\epsilon (x)e^{c\left| y\right| },\quad z=x+iy,\quad
\left| z\right| \rightarrow \infty ,  \label{condforf}
\end{equation}
where $c<2$, $\epsilon (x)\rightarrow \infty $ for $x\rightarrow \infty $,
and having no poles on the imaginary axis. However, as follows from (\ref
{qxplates}), (\ref{fnzx}), for a scalar field with $\zeta \neq \zeta _{c}$
the corresponding function has the form $f(z)=f_{m}(z)/(z^{2}b_{m}^{2}+1)$, $%
m=1,2$. In this case the subintegrand on the right of GAPF has purely
imaginary poles $\pm i/b_{m}$ for $b_{m}>0$, $m=1,2$. In analogy to the
purely imaginary zeros of $F(z)$, these poles have to be excluded from the
integral over the imaginary axis by semicircles on the right-half plane. The
integrals over these semicircles will give additional contributions
\begin{equation}
-\frac{\pi }{2\left| b_{m}\right| }\theta (b_{m})\left[ f_{m}(e^{\pi
i/2}/\left| b_{m}\right| )+f_{m}(e^{-\pi i/2}/\left| b_{m}\right| )\right] .
\label{frompoles}
\end{equation}
to the right-hand side of (\ref{sumformula}). In the case
$b_{1}=-b_{2}$ and for functions $f(z)$ having no poles on the
imaginary axis from (\ref{sumformula}) one obtains the Abel-Plana
formula in the usual form.

\end{document}